  \documentclass[preprint]{aastex}
   \usepackage{natbib}
   \usepackage{epsfig,latexsym,graphicx,subfigure,verbatim}
   \usepackage{threeparttable,hhline,longtable}
   \usepackage[T1]{fontenc}

   \newcommand{\wl}{$\lambda$}                     
   \newcommand{\kmps}{km s$^{-1}$}                
   \newcommand{\halpha}{H$\alpha$}                
   \newcommand{\rsi}{$\mathcal{R}($\ion{Si}{2}$)$}
   \newcommand{\vten}{$v_{10}($\ion{Si}{2}$)$}    
   \newcommand{\dmft}{$\Delta m_{15}$}            
   \newcommand{\vdot}{$\dot{v}$} 

\shorttitle{Spectroscopy of SN 1999ac}
\shortauthors{Garavini et al.}

\begin{document}
\title{Spectroscopic Observations and Analysis of the  Unusual Type Ia SN~1999ac}
\author{
G.~Garavini\altaffilmark { 1,2 }
G.~Aldering \altaffilmark{ 3,4 }
A.~Amadon\altaffilmark { 5 }
R.~Amanullah \altaffilmark{ 2 }
P.~Astier\altaffilmark { 2 }
C.~Balland\altaffilmark { 1,6 }
G.~Blanc\altaffilmark { 7 }
A.~Conley\altaffilmark { 3,8 }
T.~Dahl\'en \altaffilmark{ 2 }
S.~E.~Deustua \altaffilmark{ 10 }
R.~Ellis \altaffilmark{ 11 }
S.~Fabbro\altaffilmark { 12 }
V.~Fadeyev\altaffilmark { 3 }
X.~Fan\altaffilmark { 13 }
G.~Folatelli\altaffilmark { 2 }
B.~Frye\altaffilmark { 3 }
E.~L.~Gates\altaffilmark { 14 }
R.~Gibbons\altaffilmark { 15 }
G.~Goldhaber\altaffilmark { 3,8 }
B.~Goldman\altaffilmark { 16 }
A.~Goobar\altaffilmark { 2 }
D.~E.~Groom \altaffilmark{ 3 }
J.~Haissinski\altaffilmark { 17 }
D.~Hardin\altaffilmark { 1 }
I.~Hook\altaffilmark { 18 }
D.~A.~Howell\altaffilmark { 19 }
S.~Kent\altaffilmark { 20 }
A.~G.~Kim\altaffilmark { 3 }
R.~A.~Knop\altaffilmark { 15 }
M.~Kowalski\altaffilmark { 3 }
N.~Kuznetsova\altaffilmark { 3 }
B.~C.~Lee \altaffilmark{ 3 }
C.~Lidman\altaffilmark { 21 }
J.~Mendez\altaffilmark { 22,23 }
G.~J.~Miller\altaffilmark { 24,25 }
M.~Moniez \altaffilmark{ 17 }
M.~Mouchet\altaffilmark {30}
A.~Mour\~ao\altaffilmark { 12 }
H.~Newberg\altaffilmark { 26 }
S.~Nobili \altaffilmark{ 1,2 }
P.~E.~Nugent\altaffilmark { 3 }
R.~Pain\altaffilmark { 1 }
O.~Perdereau\altaffilmark { 17 }
S.~Perlmutter\altaffilmark { 3,8 }
R.~Quimby\altaffilmark { 3 }
N.~Regnault\altaffilmark { 1 }
J.~Rich\altaffilmark { 5 }
G.~T.~Richards  \altaffilmark{ 27 }
P.~Ruiz-Lapuente \altaffilmark{ 23 }
B.~E.~Schaefer\altaffilmark { 28 }
K.~Schahmaneche \altaffilmark{ 1 }
E.~Smith\altaffilmark { 15 }
A.~L.~Spadafora\altaffilmark { 3 }
V.~Stanishev\altaffilmark { 2 }
R.~C.~Thomas\altaffilmark { 3 }
N.~A.~Walton\altaffilmark { 29 }
L.~Wang\altaffilmark { 3 }
W.~M.~Wood-Vasey \altaffilmark { 3,8 }
(THE SUPERNOVA COSMOLOGY PROJECT)
}

\altaffiltext{1}{LPNHE, CNRS-IN2P3, University of Paris VI \& VII, 
Paris, France }
\altaffiltext{2}{Department of Physics, Stockholm University,  Albanova 
University Center, S-106 91 Stockholm, Sweden}
\altaffiltext{3}{E. O. Lawrence Berkeley National Laboratory, 1 
Cyclotron Rd., Berkeley, CA 94720, USA }
\altaffiltext{4}{Visiting Astronomer, Cerro Tololo Interamerican 
Observatory, National Optical Astronomy Observatory, which is operated 
by the Association of Universities for Research in Astronomy, Inc. 
(AURA) under cooperative agreement with the National Science 
Foundation.}
\altaffiltext{5}{DAPNIA-SPP, CEA Saclay, 91191 Gif-sur-Yvette, France}
\altaffiltext{6}{Universite Paris-Sud and APC, 11 Place Marcelin
Berthelot, F-75231 Paris Cedex 05}
\altaffiltext{7}{Osservatorio Astronomico di Padova, INAF, vicolo 
dell'Osservatorio 5, 35122 Padova, Italy}
\altaffiltext{8}{Department of Physics, University of California 
Berkeley, Berkeley, 94720-7300 CA, USA}
\altaffiltext{10}{American Astronomical Society,  2000 Florida Ave, NW, 
Suite 400, Washington, DC, 20009 USA.}
\altaffiltext{11}{California Institute of Technology, E. California 
Blvd, Pasadena,  CA 91125, USA}
\altaffiltext{12}{ CENTRA e Dep. de Fisica, IST, Univ. Tecnica de 
Lisboa}
\altaffiltext{13}{Steward Observatory, the University of Arizona, 
Tucson , AZ 85721}
\altaffiltext{14}{Lick Observatory, P.O. Box 85, Mount Hamilton, CA 
95140}
\altaffiltext{15}{Department of Physics and Astronomy, Vanderbilt 
University, Nashville, TN 37240, USA}
\altaffiltext{16}{Department of Astronomy, New Mexico State University, 
Dept. 4500, P.O. Box 30001, Las Cruces, NM 88011}
\altaffiltext{17}{Laboratoire de l'Acc\'elerateur Lin\'eaire, 
IN2P3-CNRS, Universit\'e Paris Sud, B.P. 34, 91898 Orsay Cedex, France}
\altaffiltext{18}{Department of Physics, University of Oxford, Nuclear 
\& Astrophysics Laboratory,  Keble Road, Oxford, OX1 3RH, UK}
\altaffiltext{19}{Department of Astronomy and Astrophysics, University 
of Toronto, 60 St. George St., Toronto, Ontario M5S 3H8, Canada}
\altaffiltext{20}{Fermi National Accelerator Laboratory, P.O. Box 500, 
Batavia, IL 60510}
\altaffiltext{21}{European Southern Observatory, Alonso de Cordova 
3107, Vitacura, Casilla 19001, Santiago 19, Chile }
\altaffiltext{22}{Isaac Newton Group, Apartado de Correos 321, 38780 
Santa Cruz de La Palma, Islas Canarias, Spain}
\altaffiltext{23}{Department of Astronomy, University of Barcelona, 
Barcelona, Spain }
\altaffiltext{24}{Department of Astronomy, San Diego State University, 
5500 Campanile Drive, San Diego, CA 92182-1221 }
\altaffiltext{25}{Department of Astronomy, University of Illinois, 1002 
West Green Street Urbana, IL 61801 }
\altaffiltext{26}{Rensselaer Polytechnic Institute, Physics Dept., 
SC1C25, Troy NY 12180, U.S.A.}
\altaffiltext{27}{Princeton University Observatory, Peyton Hall, 
Princeton, NJ 08544.}
\altaffiltext{28}{Louisiana State University, Department of Physics and 
Astronomy,
Baton Rouge, LA, 70803, USA}
\altaffiltext{29}{Institute of Astronomy, Madingley Road, Cambridge CB3 
0HA, UK }
\altaffiltext{30}{LUTH, UMR 8102 CNRS, Observatoire de Paris, Section de Meudon, F-92195 Meudon Cedex}

\begin{abstract}
We present optical spectra of the peculiar Type Ia supernova (SN Ia)
1999ac.  The data extend from $-15$ to $+42$ days with respect to
B-band maximum and reveal an event that is unusual in several
respects.  Prior to B-band maximum, the spectra resemble those of
SN~1999aa, a slowly declining event, but possess stronger \ion{Si}{2}
and \ion{Ca}{2} signatures (more characteristic of a spectroscopically
normal SN).  Spectra after B-band maximum appear more normal.  The
expansion velocities inferred from the Iron lines appear to be lower
than average; whereas, the expansion velocity inferred from Calcium H
and K are higher than average. The expansion velocities inferred from
\ion{Si}{2} are among the slowest ever observed, though  SN~1999ac
is not particularly dim. The analysis of the parameters \vten, \rsi,
\vdot, and \dmft\ further underlines the unique characteristics of
SN~1999ac.  We find convincing evidence of \ion{C}{2} \wl 6580 in the
day $-15$ spectrum with ejection velocity $v > 16,000$ \kmps, but this
signature disappears by day $-9$.  This rapid evolution at early times
highlights the importance of extremely early-time spectroscopy.
\end{abstract}

\maketitle

   \section{Introduction}

\label{sec_intro}

Our understanding on the nature of Type Ia Supernovae (SNe Ia) would
be greatly enhanced with the availability of high-quality spectra at
very early phases (the first week or so after the explosion).  At
early phases, the ejecta is denser and observations probe the faster
moving outer layers of the explosion
\citep{1992ApJ...387L..33R,1991A&A...245..114K}. The diversity in Type
Ia Supernova spectra is greatest at this time. At later phases, the
ejecta thins, we probe deeper into the slower moving layers of the
explosion and SN Ia spectra become more homogeneous.

Mapping the composition of SN Ia ejecta at high velocity leverages
direct constraints placed on explosion models.  Such constraints
include identifying the presence of ion signatures in SN spectra at
various velocity intervals, a nontrivial task since the corresponding
line profiles are often heavily blended.  In particular, the presence
of carbon at high velocity is consistent with a one-dimensional
deflagration model \citep{1984ApJ...286..644N,1985ApJ...294..619B}.
Silicon-peak elements at high velocity are more consistent with, for
example, a one-dimensional delayed-detonation model
\citep{1991A&A...245..114K}.  The signature of unburned material at a
variety of velocities discovered across a sample of SNe Ia may support
more recent three-dimensional deflagration models
\citep{2002A&A...391.1167R,2003Sci...299...77G}.

However, the task of obtaining early phase spectra presents numerous
logistical challenges.  Efficient discovery of SNe very shortly after
explosion requires both high sensitivity and reliable rejection of
false positives.  A spectrum for confirmation (that a candidate is a
SN) and classification must be obtained within a day.  A program for
intensive followup may then be triggered, requiring coordination of
limited telescope time at possibly several different sites.  Seeking
to augment the local SN Ia dataset used as calibrated candles for
cosmological distance measurement
\citep{1998Natur.391...51P,1998ApJ...493L..53G,1998ApJ...507...46S,1998AJ....116.1009R,1999ApJ...517..565P,2003ApJ...598..102K,2003ApJ...594....1T,2004ApJ...607..665R},
the Supernova Cosmology Project (SCP) organized such a program
together with several other teams \citep{2000AIPC..522...75A}
which ran during the Spring of 1999.  When SN~1999ac was reported
\citep{1999IAUC.7114....1M}, the SCP had many telescope nights
pre-scheduled for spectroscopy and photometry to observe this bright
event.  As a result, some of the earliest observed spectra of a SN Ia
were obtained.

SN~1999ac (R.A. = 16$^{h}$07$^{m}$15.0$^{s}$ Decl. =
+07$^{d}$58$\arcmin$20$\arcsec$, equinox 2000.0) was discovered and
confirmed on unfiltered observations taken on Feb. 26.5 and 27.5 UT at
23".9 east and 29".8 south of the nucleus of its Scd host galaxy, NGC
6063 \citep{1999IAUC.7114....1M}.  \citet{1999IAUC.7122....2P} reports that a
confirmation spectrum taken on Feb. 28 UT is similar to that SN~1999aa
with stronger \ion{Si}{2} \wl6355 and well defined \ion{Ca}{2} H\&K.
Optical light curves of SN~1999ac have been discussed in
\citet{Phillips:2002cg} and \citet{2003PASP..115..453L}. They report a
B-band light curve similar to that of SN~2002cx raising as fast as
SN~1991T and declining similarly to SN~1994D until two weeks after
maximum. The V-band light curve of SN~1999ac closely resembles that of
SN~2002cx until 30 days after maximum light. The similarities between
SN~1999ac and SN~2002cx break down in R and I bands.

SN~1999ac was found to be spectroscopically similar to SN~1999aa
\citep{2001ApJ...546..734L,2004AJ....128..387G}, which is considered
to be an ``intermediate'' SN Ia with properties ``between'' those of
spectroscopically normal SNe Ia and those of the spectroscopically
peculiar, bright SN~1991T \citep[for a discussion of whether or not
  this has implications for progenitor channels,
  consult][]{2001ApJ...546..734L, 2001PASP..113..169B}.  However, as
we will show in this paper, the peculiarities of SN~1999ac are not
limited to early epochs. The study of the ejecta geometry indicates
calcium lines being formed in high velocity layers and iron and
silicon lines being formed in low velocity layers.  Furthermore, when
SN~1999ac is plotted in the parameter space of \rsi, \vten, \vdot, and
\dmft\ it stands out as an outlier among other well studied SNe.

Here we present the collected spectra (our photometry of this SN will
appear in a later work) and discuss in detail the two earliest spectra
taken at days $-15$ and $-9$ with respect to the date of maximum
brightness.  The organization of this article is as follows.  In
\S\ref{sec_reduct}, a description of the reduction scheme is given.
Section \S\ref{sec_spectra} presents the spectra of SN~1999ac, and
includes some empirical analysis of the early time spectra.  Fits to
the early spectra produced with the SN spectrum synthesis code SYNOW
appear in \S\ref{sec_analysis}.  A study of the ejecta geometry is
carried out in \S\ref{ejecta} and a comparison of several spectral
indicators with those of other objects in \S\ref{parameter} and
\S\ref{sec_conclusion} concludes the article.

   \section{Data \& Reduction}

\label{sec_reduct}
\begin{figure*}
\centering 
\includegraphics[width=16cm]{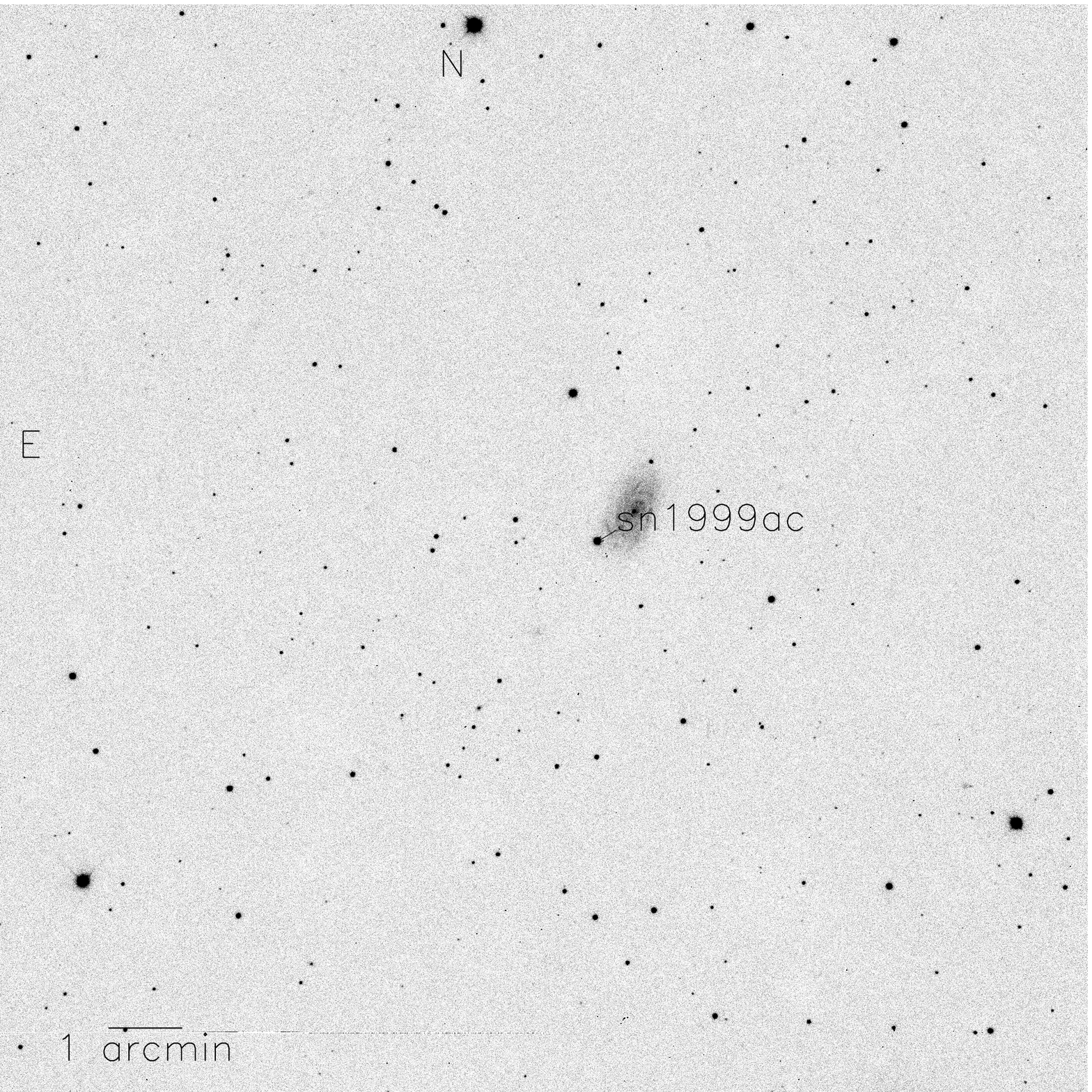}
  \caption{SN~1999ac in its host galaxy NGC 6063. R.A. =
16$^{h}$07$^{m}$15.0$^{s}$ Decl. = +07$^{d}$58$^{m}$20$^{s}$ (equinox
2000.0)}
  \label{99acfinding}
\end{figure*}

Figure \ref{99acfinding} shows the position of SN~1999ac in its host
galaxy NGC 6063, an Scd galaxy with a recession velocity of 2848 \kmps\
as determined from narrow \halpha\ emission \citep{1998A&AS..130..333T}.  All
spectra presented in this paper have been shifted to rest frame using
this recession velocity.

The data set  
consists of 14 optical spectra extending from day $-15$ to
day $+42$ (in this work all phases are expressed with respect to B-band
maximum).  In most cases, the spectra were acquired using different
instrumental settings for the blue and red parts of the spectrum to
avoid possible second-order contamination.  Hence, the fully reduced 14
spectra are the combinations of both blue and red parts.  The observation 
log appears in Table \ref{tabdata}.

The data were reduced using standard IRAF routines.  The
two-dimensional images were bias-subtracted and flat-fielded.  The sky
background was fitted, subtracted, and extracted for systematics
checks on the wavelength calibration.  Wavelength and flux calibration
were applied to the one-dimensional extracted spectra using
calibration observations taken with the same instrumental setting and
during the same night as science observations.  The accuracy of the
wavelength calibration was checked against the extracted sky spectra
and generally found to agree to within 2~\AA.  Synthetic
spectrophotometry was computed to check the overall accuracy of the
flux calibration against the publicly available photometric data
\citep{jhathesis} and found to agree within the quoted statistical
uncertainties.  An atmospheric extinction correction was applied via
tabulated extinction coefficients for each telescope used.

Galactic reddening in the
direction of SN~1999ac was estimated following
\citet{1998ApJ...500..525S}.
A value of $E(B-V) = 0.046$ mag was found. For host galaxy extinction,   
\citet{Phillips:2002cg} found that SN 1999ac's photometric evolution does not
follow the Lira-Phillips relation \citep{1999AJ....118.1766P}, making
problematic 
an estimate of host galaxy extinction with this technique.
\citet{2001BAAS...33.1207.} tentatively derive a total extinction to
SN~1999ac of $A_V = 0.51$ mag. 
All spectra of SN~1999ac were corrected for Galactic extinction 
assuming $R_V = 3.1$, i.e. $A_V = 0.14$
\citep{1989ApJ...345..245C} but no attempt was made to correct for host galaxy
extinction correction because of the mentioned uncertainties in its
estimate.  This has no impact on the analysis presented, since we
focus our analysis on line profiles, which cover wavelength scales smaller
than those affected by reddening.  For a given line profile,
the effect of reddening can be considered as a global flux offset.

The amount of host galaxy light contamination was estimated by
$\chi^2$-fit of the data with a SN spectrum template. It was found to
be negligible at all epochs, and hence no subtraction of host galaxy
light beyond that included in the sky background reference regions was
performed.

\begin{figure*}
\centering
  \includegraphics[width=16cm]{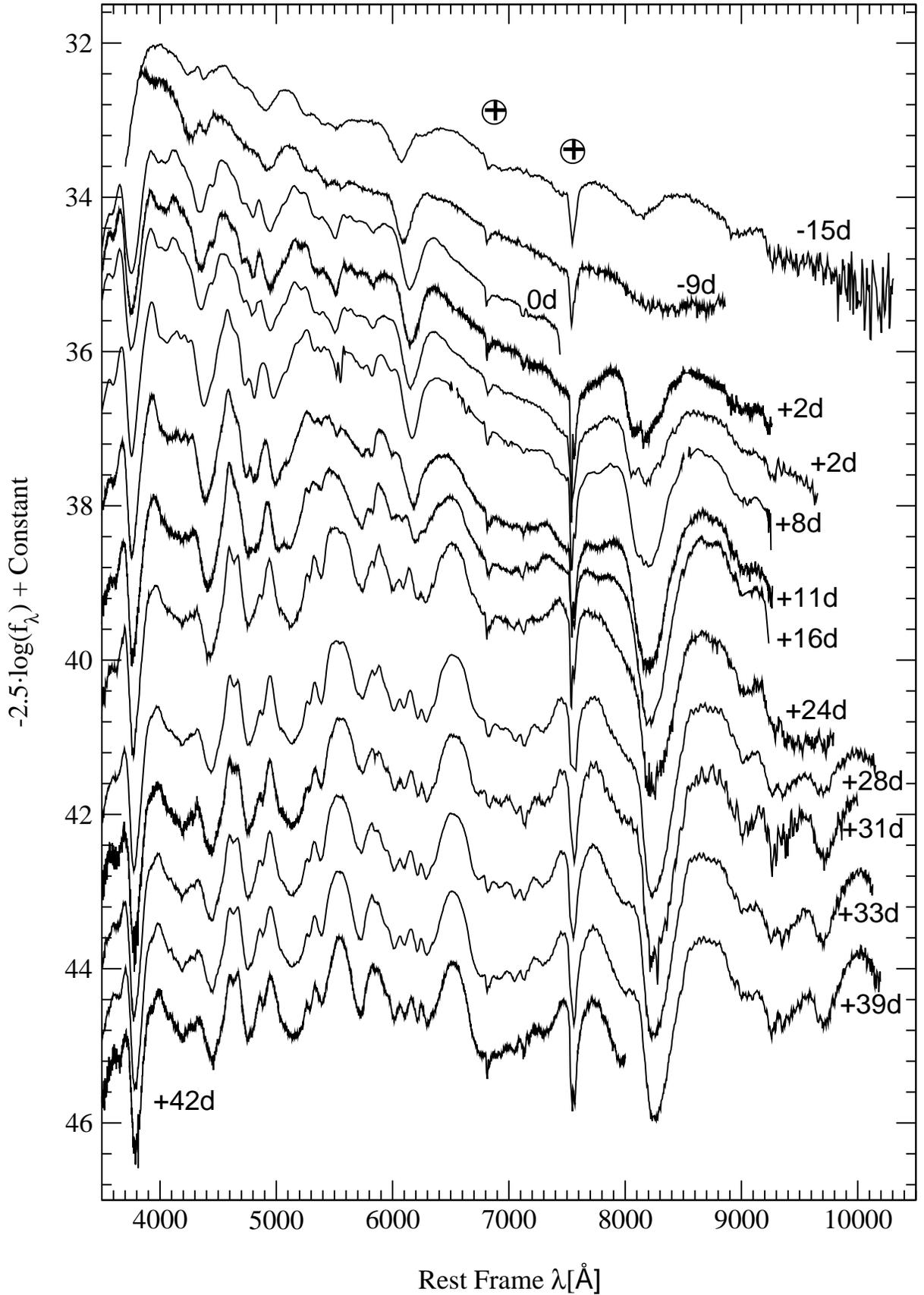}
  \caption{SN~1999ac spectral evolution. Epochs referred to B-band
    maximum light. The $\oplus$ symbol marks sharp atmospheric
    absorption features.}
  \label{ac_evo}
\end{figure*}

\begin{figure*}
\centering\includegraphics[width=16cm]{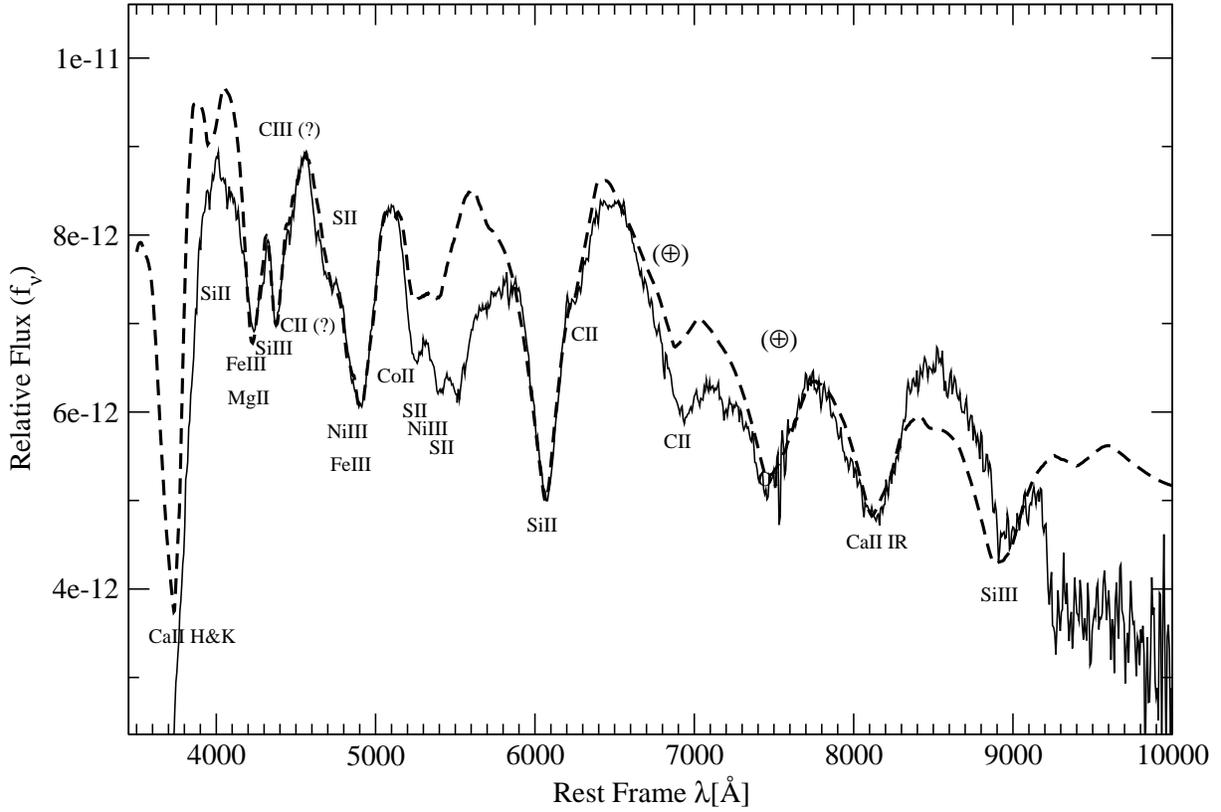}
  \caption{Synthetic spectra compared with SN~1999ac spectrum for
  $-$15 days. Dashed line: best match synthetic spectrum. Solid line:
  data. SYNOW parameters used are presented in table
  \ref{table-15}. Ions responsible for features in the synthetic
  spectrum are marked.  Where a telluric feature has been removed,
  the spectrum is marked with an Earth symbol in parenthesis.}
  \label{synow-15}
\end{figure*}

Telluric corrections are applied to spectra used for synthetic
spectrum analysis, using calibration spectra from spectrophotometric
standard stars in the vicinity of SN 1999ac.  The wavelength regions
with telluric corrections are marked in the figures, and their effect
can be seen by comparing Figure \ref{ac_evo} with Figure
\ref{synow-15}, which show the spectra after telluric
correction.

   \section{Spectra}
\label{sec_spectra}

The timespan and sampling frequency of the data set permits study of the
spectroscopic evolution of SN~1999ac from very soon after explosion to
seven weeks beyond B-band maximum light.  Figure \ref{ac_evo} presents
the 14 fully reduced spectra, along with phases of observation for each.
 The top spectrum, taken at day $-15$, is one of the earliest ever
obtained of a SN Ia.

The two earliest spectra are relatively featureless, though they clearly
show the ion signatures of a SN Ia.  The more obvious absorption
features  are at 4900, 6100, and 8100~\AA; due to \ion{Fe}{3} blends,
the \ion{Si}{2} \wl 6355 blend, and the \ion{Ca}{2} infrared (IR)
triplet (respectively).  A weak absorption at 4200~\AA\ is probably also
due to \ion{Fe}{3}, while \ion{Si}{3} is responsible for a weak
absorption at 4400~\AA.  Two very weak notches at 5200 and 5500~\AA\
hint at the presence of the \ion{S}{2} ``W'' feature.  A small
depression just redward of the \ion{Si}{2} 6100~\AA\ absorption could be
due to \ion{C}{2}.  We return to this issue later in this article.

The absorptions strengthen up to about day $+8$.  The
\ion{Si}{2} \wl 6355 absorption shifts noticeably to the red during this
time.  After day $+8$, the \ion{Ca}{2} IR triplet profile morphology
remains roughly constant.  The \ion{S}{2} ``W'' feature strengthens at
maximum light, but shortly afterward disappears as it is replaced by
\ion{Fe}{2} and \ion{Na}{1} lines.  

The absorption at 6100~\AA, due to \ion{Si}{2} \wl 6355, changes shape
between days $+8$ and $+24$ as \ion{Fe}{2} lines strengthen and
obliterate it through line blending.  By day $+24$, four robust minima
have replaced the 6100~\AA\ absorption.  Furthermore, the ostensible
\ion{Si}{2} emission peak appears to shift redward during this
interval.  The simplest explanation for this behavior is not a real
shift {\it per se}, but rather an effect of blending, as the bluest
part of the emission feature is overcome by the reddest absorption
notch.  It is interesting to note that \ion{Fe}{2} lines begin
contaminating the \ion{Si}{2} feature as early as day $+11$, somewhat
earlier than usual in SNe Ia.

From day $+24$ onward, however, the spectrum basically does not evolve.
The transition to iron-peak species dominance is complete just three
weeks after maximum light, as numerous iron-peak lines dominate the
spectrum from the ultraviolet to the near IR, excluding the \ion{Ca}{2} H\&K and
IR triplet features.

   \subsection{Early-Time Comparisons}

Generally, the post-maximum spectroscopic evolution of SN~1999ac
resembles that of a normal SN Ia.  The early-time spectra (days $-15$ and
$-9$) present the unusual opportunity of comparison with 
the small set of supernovae with similarly early spectra.  

\begin{figure*}
\centering\includegraphics[width=16cm]{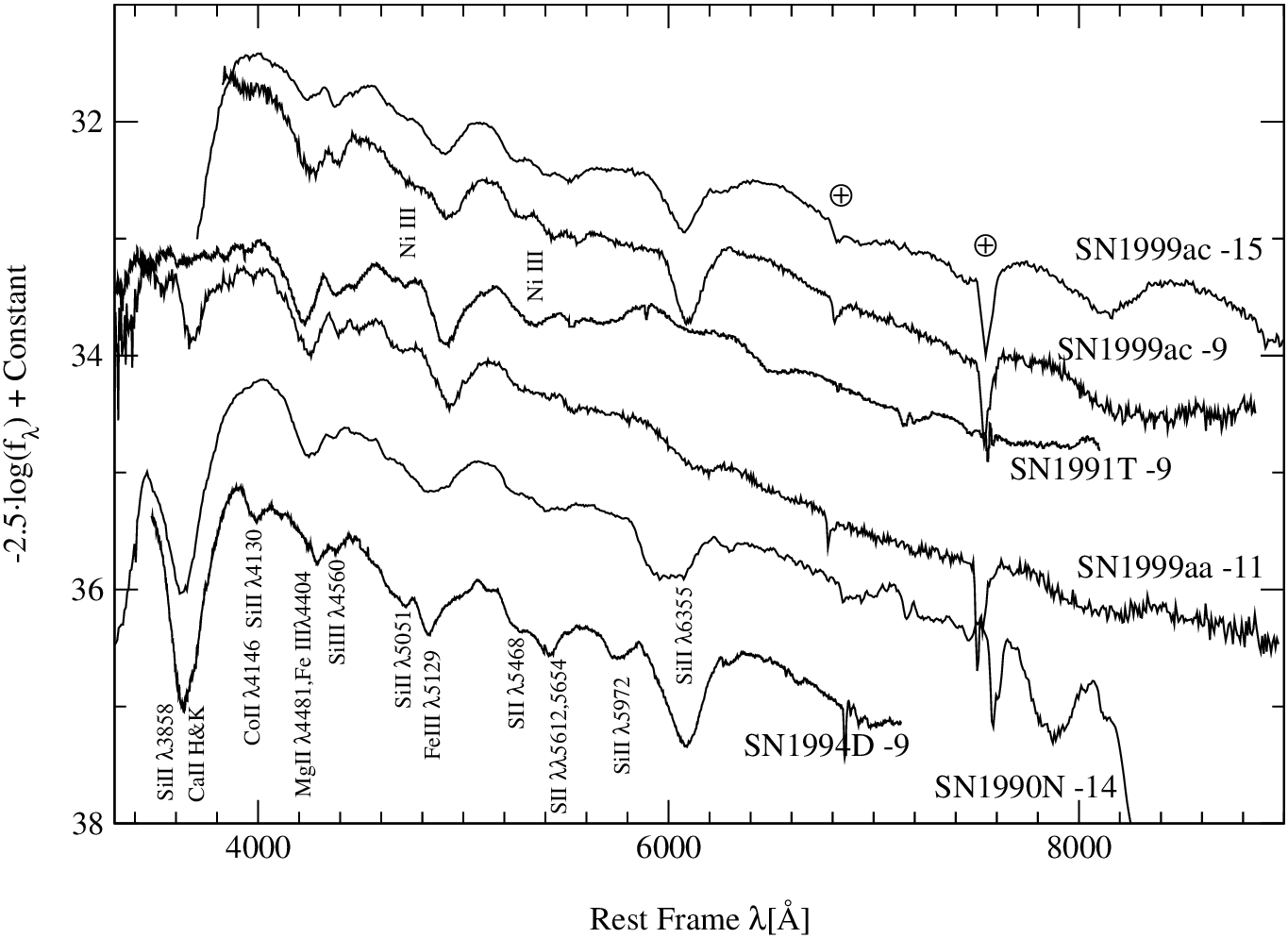}
   \caption{The $-$15 and $-$9 days spectra of SN~1999ac together with
  those of SN~1999aa, SN~1991T, SN~1990N and SN~1994D respectively
  from
  \citet{2004AJ....128..387G,1992ApJ...384L..15F,1991ApJ...371L..23L,1996MNRAS.278..111P}. Epochs
  are quoted in the labels. Line identification are taken as in
  \citet{1999AJ....117.2709L,2001PASP..113.1178L,1999MNRAS.304...67F,1996MNRAS.278..111P,1995A&A...297..509M,1993ApJ...415..589K,1992ApJ...397..304J}}.
  \label{-15_comparison}
\end{figure*}

In Figure \ref{-15_comparison}, the day $-15$ spectrum is presented
along with early-time spectra of SNe 1991T \citep{1992ApJ...384L..15F}, 1999aa
\citep{2004AJ....128..387G}, 1990N \citep{1991ApJ...371L..23L}, and 1994D
\citep{1996MNRAS.278..111P}.  Compared to the spectroscopically peculiar SNe~1999aa
and 1991T, SN~1999ac exhibits stronger \ion{Ca}{2} and \ion{Si}{2} \wl
6355 absorptions.  On the other hand, its \ion{Fe}{3} 4400, 4100~\AA\
and \ion{Si}{3} \wl 4560 features are weaker.

A small flux depression in the day $-15$ spectrum is clearly visible to
the red of the \ion{Si}{2} 6100~\AA\ feature, echoing one visible in the
spectrum of SN~1990N.  Though made uncertain by the presence of the
telluric absorption at 6900~\AA, another feature at 7000~\AA\ could be
common to both spectra as well.  The first feature has been identified
as \ion{C}{2} \wl 6580 in SN~1990N \citep{2001MNRAS.321..341M}, in SN~1994D
\citep{1999ApJ...525..881H}, in SN~1998aq \citep{2003AJ....126.1489B}, and in SN~1999aa
\citep{2004AJ....128..387G}.  The second feature could be due to another line
from the same ion, \ion{C}{2} \wl 7234.

Overall, the $-15$ day spectrum resembles that of SN~1990N more than
it does SN~1994D, with its more obvious (and characteristic)
\ion{S}{2} and \ion{Fe}{2} blends.  In summary, at day $-15$,
SN~1999ac possess spectroscopic characteristics common to both
slow-declining SNe like SN~1991T and spectroscopically normal ones
like SN~1990N at a similar phase.

The $-9$ day spectrum of SN~1999ac (Figure \ref{-15_comparison}) does
not extend to the \ion{Ca}{2} H\&K absorption.  The two \ion{Fe}{3}
features are still weaker than in SN~1991T and possess rounded minima.
The contribution from \ion{Fe}{2} in the region between 4000 and 5000
\AA\ appears weaker than in SN~1994D and in SN~1990N.  The absorption
feature present on the red edge of this line and that around 7000~\AA\
at day $-15$ are no longer evident.

   \section{Synthetic Spectra}

\label{sec_analysis}

One of the more interesting ion signatures suggested in the previous
section is that of \ion{C}{2}.  The presence of carbon lines and their
ejection velocities pose potential constraints on hydrodynamical explosion
models.  Carbon at high velocity is consistent with both
one-dimensional \citep{1984ApJ...286..644N} and three-dimensional
\citep{2002A&A...391.1167R,2003Sci...299...77G} deflagration models.
An apparently generic result from the newer models is the mixing of
``fuel'' (carbon and oxygen) and ``ashes'' (products of
nucleosynthesis) at all velocities.  Hence, the detection of carbon at
low velocity would support the three-dimensional deflagrations.

To further explore the \ion{C}{2} signature and its behavior at early
times, in this section we compare synthetic spectra to those observed
before maximum light.  Clearly, the ideal approach would be to  {\it
invert} a SN spectrum in some way to yield a composition model.
Unfortunately, the SN atmosphere problem is an ill-posed inverse
problem.  Instead, one of two approaches must be adopted.  

One approach is {\it detailed} analysis, where the goal is to include
all relevant transfer physics (nonlocal thermodynamic equilibrium rates,
relativity, time-dependence, energy from radioactive decays, etc.) and
numerically simulate the emergent spectrum of a given hydrodynamical
model.  This approach is appropriate for validating hydrodynamical
models, or suggesting adjustments to such models in the future.  Though
powerful, detailed analysis codes \citep[for example, the
general-purpose PHOENIX code,][]{Hauschildt1999} consume months of
computer time for a single calculation.

Another, complimentary approach is {\it direct} analysis.  The goal of
direct analysis is more empirical; to constrain the presence or absence
of ions in a spectrum and the ejection velocities of their parent atoms.
The task is generally nontrivial, since SN spectra consist of many
blends of lines which cannot be treated simplistically.  Still,
approximate techniques are used to make the process fast and iterative,
but the results are powerful.  Constraints from direct analysis are of
use to both detailed modelers and explosion modelers.  In the latter
case, the results of direct analysis can rule out many hydrodynamical
models (in principle) before the extensive simulation calculations 
are performed. 

SYNOW \citep{2000PhDT.........6F} is a direct analysis code that
generates spectra based on a simple, conceptual model of a SN
appropriate during the first few weeks to months after explosion.
This model consists of a blackbody-emitting, sharply defined
photosphere surrounded by an extended line-forming, pure scattering
atmosphere.  The entire envelope is assumed to be homologously
expanding.  Line transfer is treated using the Sobolev method
\citep{1960mes..book.....S,1970MNRAS.149..111C,1990sjws.conf..149J} so
line opacity is parameterized in terms of Sobolev optical depth.
Which ions are used in the calculation is determined by experience,
guided by the SN ion signatures atlas of \citet{1999ApJS..121..233H}.
For each ion introduced, Sobolev optical depth as a function of
velocity for a ``reference line'' (usually a strong optical line) is
specified.  Optical depths in other lines of the ion are set assuming
Boltzmann excitation of the levels at temperature $T_{exc}$.

The parameters $v_{phot}$ and $T_{bb}$ set the velocity and blackbody
continuum temperature of the photosphere, respectively.  For each ion,
the optical depth $\tau$ at the minimum ejection velocity $v_{min}$ is
specified.  Optical depth scales exponentially with velocity according
to the $e$-folding velocity parameter $v_{e}$, and is considered to be
zero for velocity greater than $v_{max}$.  If $v_{min} > v_{phot}$ for
an ion, we refer to the ion as ``detached.''

A sharply defined, blackbody-emitting photosphere clearly cannot serve
as a perfect substitute for the processes of continuum formation in a SN
atmosphere.  Hence, the synthesized continuum level may systematically 
differ from that observed in some wavelength regions.  Generally, a good
fit to the blue continuum results in a brighter red synthetic continuum.
This effect has little to no bearing on line identifications or velocity
inferences.

   \subsection*{Day $-15$}

Figure \ref{synow-15} is a comparison of a synthetic spectrum to the
observed spectrum at day $-15$.  Choosing $T_{bb} = 11,200$ K reproduces
a satisfactory reproduction of the overall continuum shape.  We also
find that the choice of $v_{phot} = 13,000$ \kmps\ assists in producing
reasonable line profiles when other factors are taken into account for
each individual ion.

The \ion{Si}{2} optical depth profiles are detached, with $v_{min} =
14,200$ \kmps.  This detachment moves the synthetic \ion{Si}{2} \wl 6355
absorption higher in velocity space so that it matches the observed
feature.  It also flattens the corresponding emission feature to improve
the agreement.  We regard the presence of \ion{Si}{2} in this spectrum
as definite.

\ion{Si}{3} optical depth must be capped at $v_{max} = 17,000$
\kmps\ to match the feature usually associated with it at 4400~\AA.
Interestingly, a concomitant synthetic feature appears to reproduce a
feature in the near-infrared around 9000~\AA.  Though the fit is not
perfect here, we regard the presence of \ion{Si}{3} in this spectrum
as definite.

The velocity range in which we introduce \ion{Ca}{2} is mainly
constrained by the \ion{Ca}{2} IR triplet since the H\&K component is
missing from the data.  Nevertheless, the agreement with the observed
\ion{Ca}{2} IR triplet and the falling edge of the observable H\&K
signature is convincing.  The presence of \ion{Ca}{2} in this spectrum
is definite.

Introducing \ion{Fe}{3} provides a match to two observed features, one
at 4200~\AA\ and another at 4900~\AA.  Like \ion{Si}{2}, the
corresponding optical depth profile must be detached (in this case to
$v_{min} = 14,500$ \kmps) to reproduce the features.  Adding some
\ion{Mg}{2} improves the fit to the absorption at 4200~\AA.
The presence of \ion{Fe}{3} is definite.

As previously mentioned, the characteristic \ion{S}{2} feature does
not yet appear to be fully developed at phase $-15$ days.  Including
some \ion{S}{2} optical depth produces an absorption blend around 5300
\AA.  Generally, however, fits to the \ion{S}{2} blend region using
SYNOW do not reproduce the observed features in most cases. Our fit in
this region is not optimal, nevertheless, we regard the presence of
\ion{S}{2} in the day $-15$ spectrum as probable.

It is interesting that including some \ion{Ni}{3} to the synthetic
spectrum helps to account for the flux deficit to the blue of the
\ion{Fe}{3} 4900~\AA\ feature and near 5300~\AA.  The appearance of this
ion in a spectrum is somewhat unusual, nevertheless the improvement
cannot be discounted.  Conservately, however, we regard the presence of
\ion{Ni}{3} as possible.

\begin{figure*}
\centering\includegraphics[width=16cm]{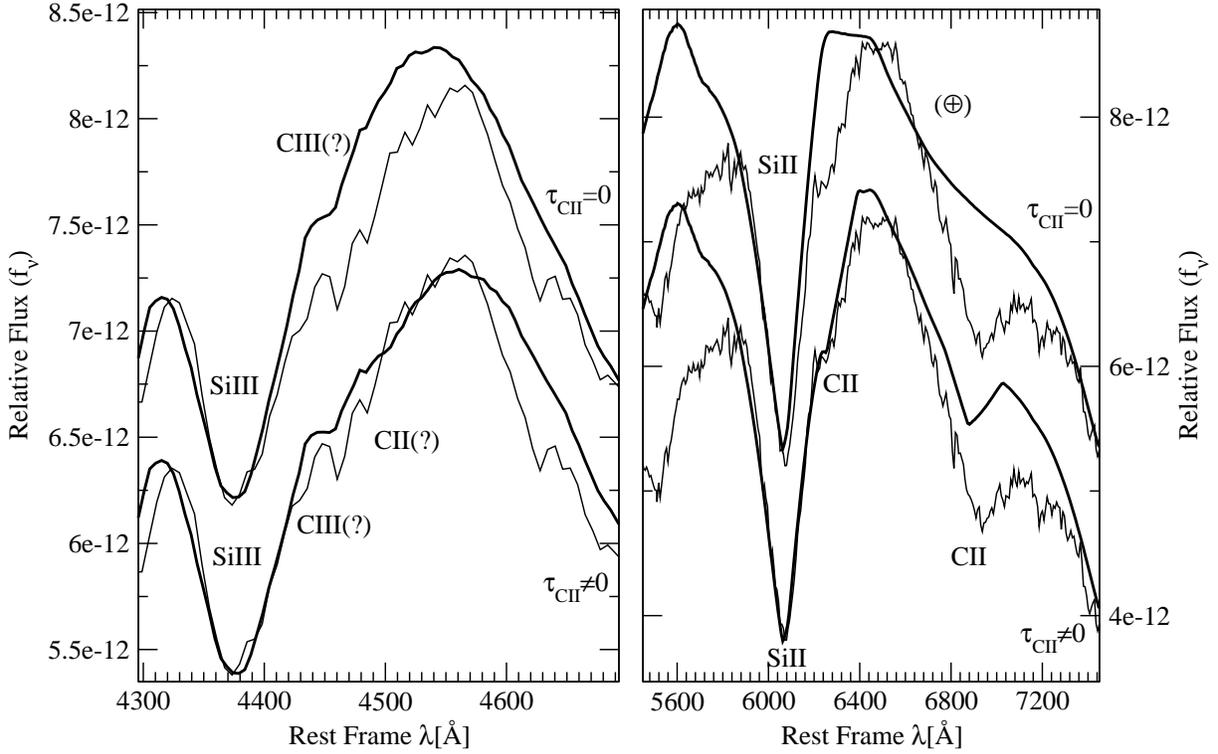}
  \caption{Synthetic spectra (heavy solid line) 
compared with SN~1999ac spectrum (light solid line) for 
  $-$15 days in the 4500~\AA\ (left panel) and 6150~\AA\ (right panel)
  region. First model from the top: $\tau_{\rm CII}=0$; 
Second model from the top: $\tau_{\rm CII}\neq0$. 
SYNOW parameters used are presented in table
  \ref{table-15}. Ions responsible for features in the synthetic
  spectrum are marked.}
  \label{CII-15}
\end{figure*}

The weak absorption at 6300~\AA\ is well matched by a detached
\ion{C}{2} (v$_{min}=16,000$~km s$^{-1}$) that may also contribute to a
feature near 4500\AA\ and perhaps at 7000~\AA.  The wavelength regions
where \ion{C}{2} makes its contribution are highlighted in Figure
\ref{CII-15} showing the effect of the presence of this ion on the
synthetic spectrum.  The good matching of the absorption feature at
6300~\AA\ and the possible contribution near 4500~\AA\ makes the
identification of \ion{C}{2} definite in the day $-15$ spectrum.

We have considered \ion{C}{3} to match the small notch on the red side
of the \ion{Si}{3} feature near 4500~\AA.  However, we find the evidence
for \ion{C}{3} less convincing than that for \ion{C}{2}.  The small
optical depth used to generate the \ion{C}{3} feature prevents the
appearance of other, weaker \ion{C}{3} lines in the synthetic
spectrum.  Hence, the only sign of \ion{C}{3} (an unusual
identification) would be this line by itself.  Without concomitant
\ion{C}{3} lines, we consider this ion to be only a remote possibility.
The feature is probably not due to H$\beta$, since the spectrum shows no
other signs of hydrogen.

   \subsection*{Day $-9$}

\label{-9}
\begin{figure*}
\centering \includegraphics[width=16cm]{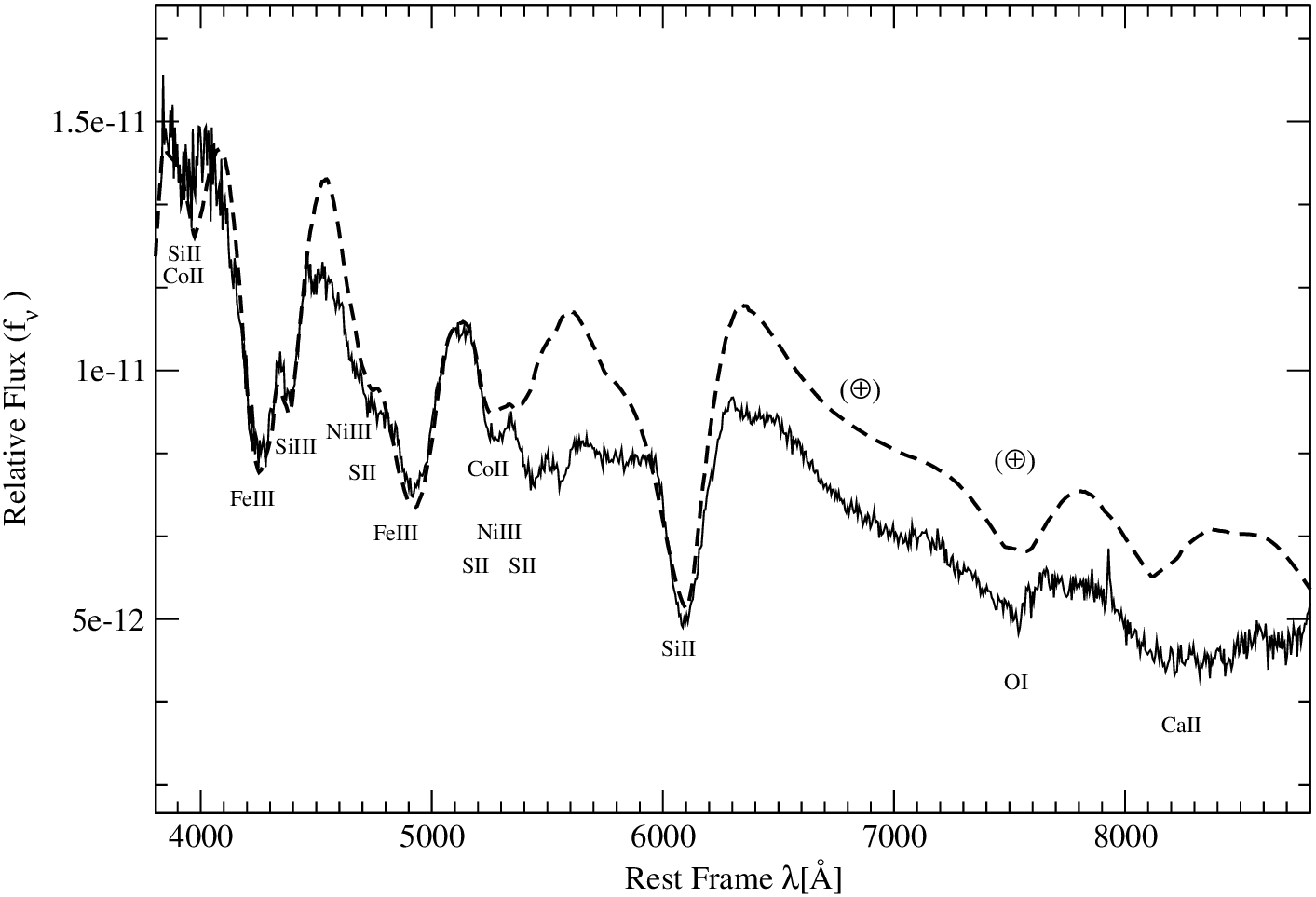}
  \caption{Synthetic spectra compared with SN~1999ac spectrum for 
  $-$9.days Dashed line: best match synthetic spectrum. Solid line:
  data. SYNOW parameters used are presented in table
  \ref{table-9}. Ions responsible for features in the synthetic
  spectrum are marked.  Where a telluric feature has been removed,
  the spectrum is marked with an Earth symbol in parenthesis.}
  \label{synow-9}
\end{figure*}

Figure \ref{synow-9} presents the synthetic spectrum produced for
comparison with the day $-9$ spectrum.  The fit parameters used appear
in Table \ref{table-9}.  The photospheric velocity has been lowered to
$v_{phot} = 11,800$ \kmps, and the blackbody temperature increased to
$T_{bb} = 13,800$ K. The increased temperature should not be
interpreted as a peculiarity since the blackbody continuum used by
SYNOW has to be regarded only as an approximation with little physical
meaning (see section \ref{sec_analysis} for details.)  For the most
part, the same ions are used for this fit as in the previous one.
Note that at this phase, SYNOW is overestimating the continuum through
the \ion{Si}{2} 6100~\AA\ absorption and redward.  However, this
offset has no effect on line identifications or ejection velocity
interval measurements.

Ions definitely present in the spectrum include \ion{Si}{2} (though no
longer detached), \ion{Si}{3} and \ion{Fe}{3}.  Again the fit to
\ion{S}{2} is problematic, but its identification based on the two
notches redward of 5200~\AA\ is not unreasonable.  \ion{Ni}{3} seems to
remove some excess flux again near 4700~\AA, but it no longer is
sufficient to account for all of the flux deficit.

Most interestingly, it appears that between day $-15$ and day $-9$, the
\ion{C}{2} signature has disappeared.  A small notch near 4500~\AA\ could
be due to a small amount of \ion{C}{3} at photospheric velocities, but
adding \ion{C}{3} optical depth only enhances the excess flux to the red
of this feature.  For this reason, we consider the presence of
\ion{C}{3} in this spectrum as unlikely, though no alternative has been
identified.

\section{Ejecta geometry} 
\label{ejecta}
We have seen in previous sections tentative evidence of C II lines
moving at velocities above 20,000 km/sec. The peculiarity of SN~1999ac
is not limited to the high velocity ejecta. Distinctive
characteristics are noticed in many other absorption features pointing
out an overall unusual ejecta geometry.  This can be investigated, in
first approximation, by looking at the line profile of spectral
features.  The characteristic line profile during the first several
weeks after outburst is a P-Cygni profile.  We here briefly review the
relevant features of the P-Cygni line profile before presenting
measurements based on it.

A P-Cygni profile arises from a configuration consisting of an
extended, expanding line-forming region surrounding an optically thick
core. Consider such a configuration, observed in the frame coinciding
with the rest frame of the core's center.  Material in front of the
core (as observed) moves toward the observer, and scatters radiation
out of the line of sight, resulting in an absorption feature
blueshifted with respect to the line rest wavelength.  Material not in
front of the core (and expanding away from it) scatters radiation into
the observer line of sight, resulting in an emission feature peaked at
the line rest wavelength.  Together, the blueshifted absorption and
rest-wavelength centered emission features form the P-Cygni profile.

The morphology of a P-Cygni absorption component provides an estimate of
the velocity interval in which the originating line forms. The blue
edge of the absorption feature forms at the highest ejection velocities
where the line is optically thick. In practice, the minimum of an
absorption profile is used to derive a characteristic ejection velocity
describing where in velocity space a line forms.  

The strength of a line can easily influence its shape.  A weak line
(with smaller line opacity) produces a sharp, robust minimum that can be
measured with little ambiguity.  Conversely, a strong line produces a
more rounded absorption feature, and as line strength increases, the
position of the absorption minimum shifts to the blue.  Hence, a
velocity measured from such a minimum is not necessarily representative
of the minimum ejection velocity of a line.

Two lines from the same parent ion, or {\it concomitant lines}, that are
closely spaced in wavelength (as in a doublet) may generally be treated
as a single line for the purposes of velocity estimation.  On the other
hand, features from other ions blending with a given line present a
problem, as they can easily shift the position of the line minimum to
the blue or red.  Therefore, care must be exercised when making
inferences from absorption features that may or may not be suffering
blending effects.

\subsection{Ejecta geometry comparison}

In the following paragraphs we compare the velocity field of some of
the characteristic features of SN~1999ac with those of other SNe known
for having a peculiar ejecta geometry, namely SN~2002cx
\citep{2003PASP..115..453L}, SN~1999aa \citep{2004AJ....128..387G} and
SN~2002bo \citep{2004MNRAS.348..261B}, and with that of SN~1994D
\citep{1996MNRAS.278..111P} which we here take as a prototype of
normal supernova. SN~2002cx is a well studied under-luminous
supernova with normal B-band light curve decline rate, spectral
signatures similar to SN~1991T but with spectral lines with low
expansion velocities. SN~2002bo had a normal decline rate but showed
spectra with higher than average expansion velocities similar to
SN~1984A \citep{1989A&A...220...83B}. Finally, SN~1999aa was found to
have a slow light curve decline rate and very weak \ion{Si}{2}
absorption features with expansion velocities constant in time. Also,
its spectra had similarities with SN~1991T at early epochs, but
rapidly changed toward normal looking spectra just before maximum
light.

\subsubsection{Pre-maximum spectra}

\begin{figure*}
\centering \includegraphics[width=12cm]{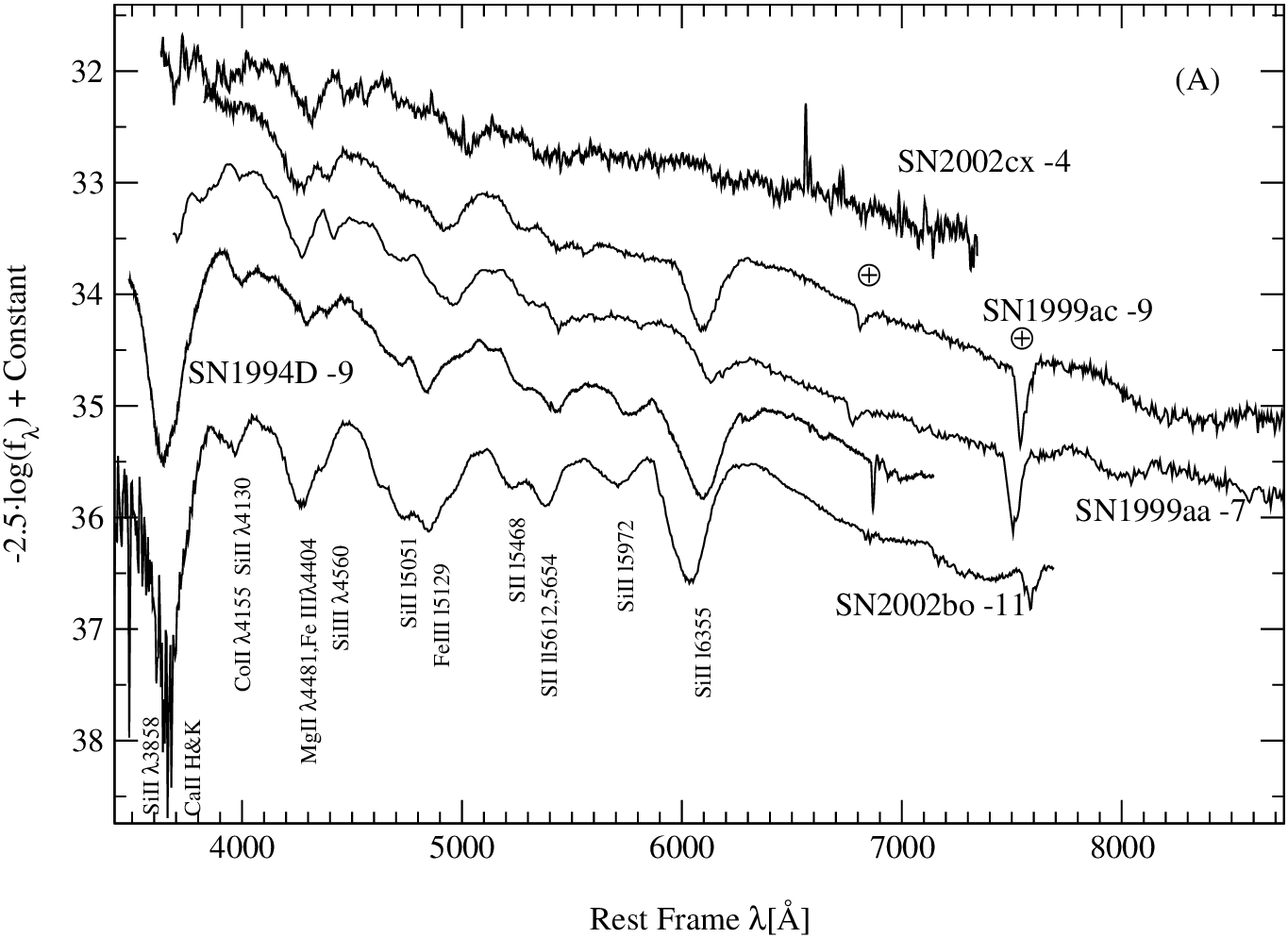}\\
\includegraphics[width=4cm]{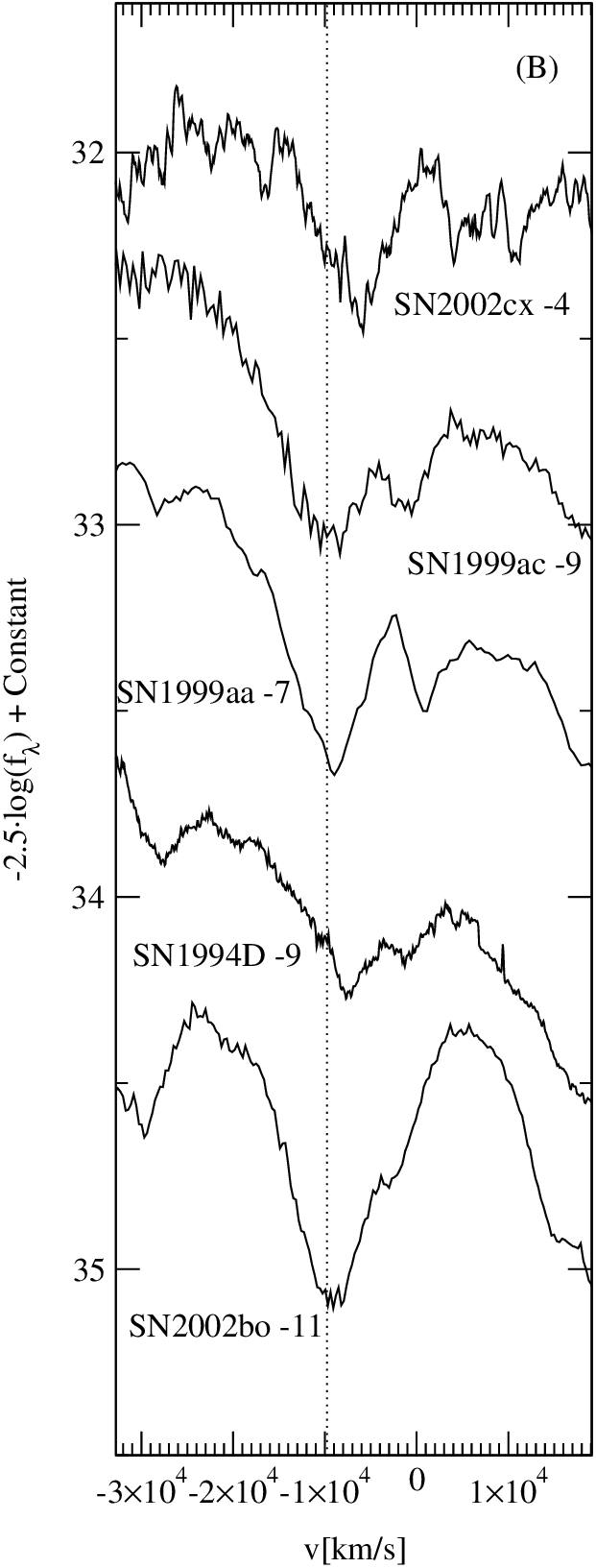}
\includegraphics[width=4cm]{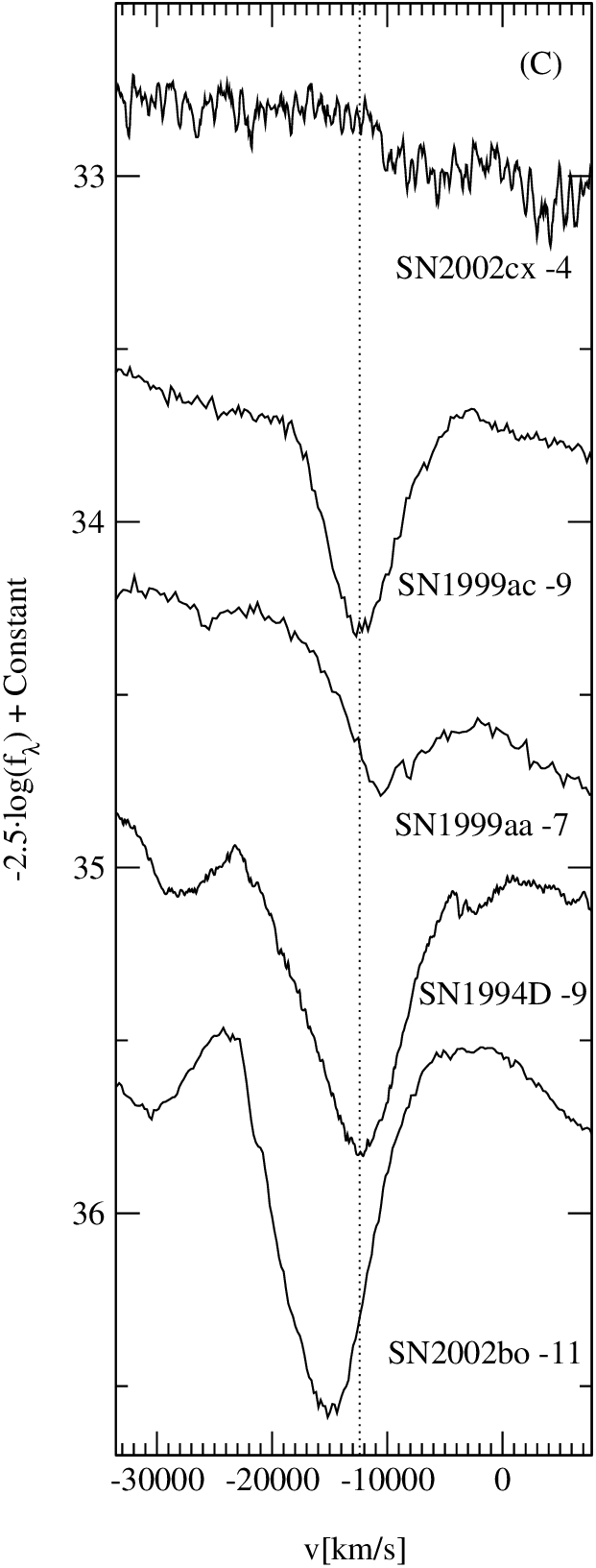}
  \caption{ {\it Panel A}: The $-$9 day spectrum of SN~1999ac
    together with those of SN~1999aa, SN~2002cx and SN~1994D. Epochs
    are quoted in the labels.  Line identifications are taken as in
    \citet{1999AJ....117.2709L,2001PASP..113.1178L,1999MNRAS.304...67F,1996MNRAS.278..111P,1995A&A...297..509M,1993ApJ...415..589K,1992ApJ...397..304J}. {\it
      Panel B}: Comparison of the same SNe in the region of
    \ion{Fe}{3}~$\lambda$4404 in velocity space. {\it Panel C}:
    Comparison of the same SNe in the region of
    \ion{Si}{2}~$\lambda$6355 in velocity space. In {\it Panel B and
      C} vertical dotted lines indicate the position of the feature
    minima for SN~1999ac.}
\label{comp99ac-9_new}
\end{figure*}

The comparison of the spectrum of SN~1999ac at day $-9$ with those of
the SNe mentioned above is shown in Figure \ref{comp99ac-9_new}, {\it
Panel A}. The most noticeable differences lie in the region of
\ion{Fe}{3} and \ion{Si}{2} lines. To investigate such differences, 
we show, in {\it Panel B} and {\it Panel C}, the comparison in velocity
space of the absorption features in the wavelength region respectively
around 4250~\AA\ and 6150~\AA.

In early spectra the absorption visible at 4250~\AA\ is generally due
mostly to \ion{Fe}{3}~\wl4404 but a contribution of \ion{Mg}{2}~\wl4481 is
expected and can vary from object to object. Normal SNe have a
stronger \ion{Mg}{2} component whereas SN~1991T-like SNe are dominated
by \ion{Fe}{3}.  Among the SNe we compare, the \ion{Mg}{2}
contamination is probably more important for SN~1994D and
SN~2002bo. SN~1999ac appears to have approximately the same velocity
distribution as SN~2002bo and as SN~1999aa. Both SN~1994D and
SN~2002cx have lower velocities than our object. The actual ejecta
geometry is difficult to disentangle from this comparison because of
the uneven contamination of \ion{Mg}{2} among the different
objects. However, based on the analysis we performed in 
\S\ref{-9}, this absorption in SN~1999ac appears to be the result of
both \ion{Mg}{2} and a dominant component of low velocity
\ion{Fe}{3}. We will come back later to this point for more
discussion.

In {\it Panel C} we compare the line profile in velocity space of the
absorption feature at around 6150~\AA. This line is due to
\ion{Si}{2}~\wl6355 and it is usually the one suffering the least
blending among all the supernova optical spectral features.  It is thus
considered the simplest to use for expansion velocity studies. At this
epoch SN~1999ac appears to have its line minimum at the same velocity
as SN~1994D but with a lower velocity for its blue edge. This is consistent
with the optical depth of \ion{Si}{2} being weak in the outermost
layer. We will come back to this point later in the
analysis. SN~2002bo and SN~2002cx show respectively a faster and
slower \ion{Si}{2} layer.

\subsubsection{Spectra at maximum}

\begin{figure*}
\centering\includegraphics[width=12cm]{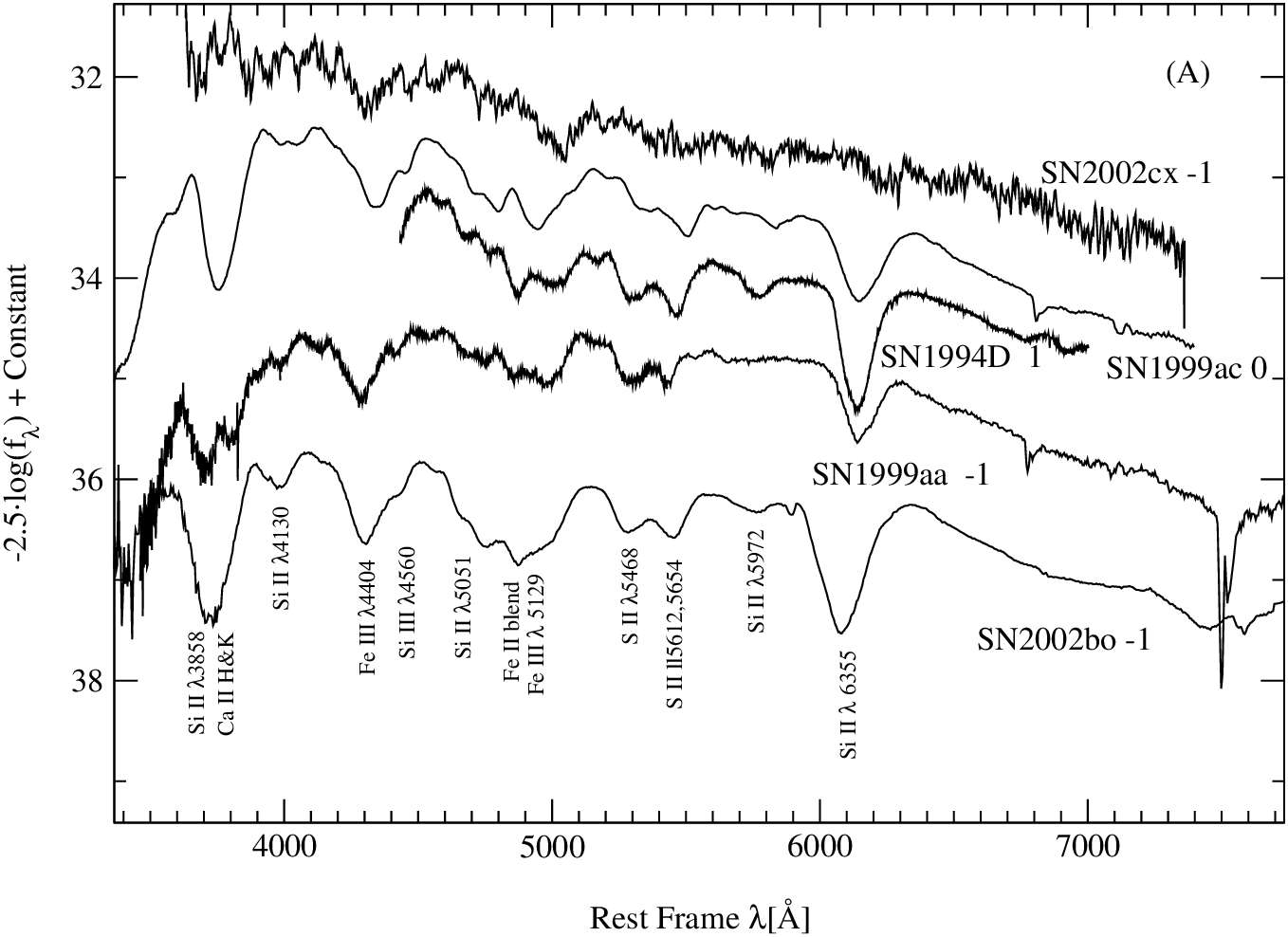}\\
\includegraphics[width=4cm]{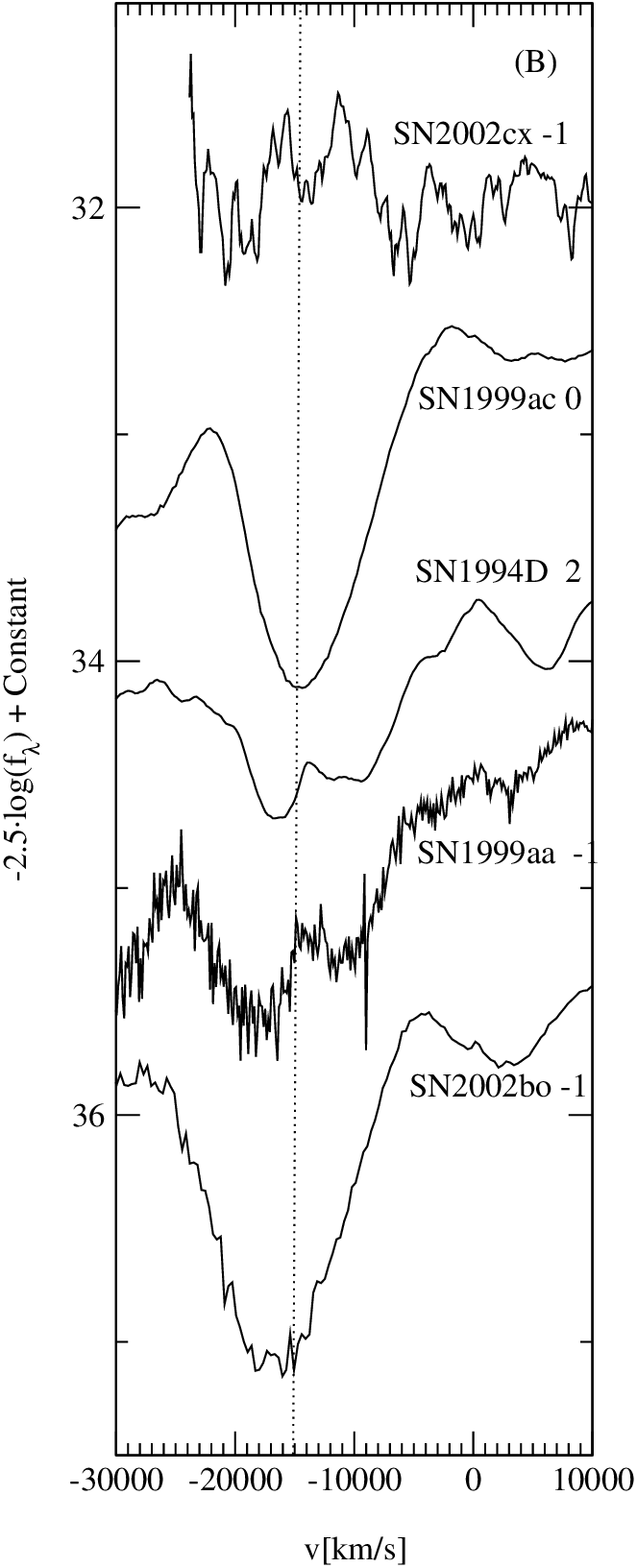}
\includegraphics[width=4cm]{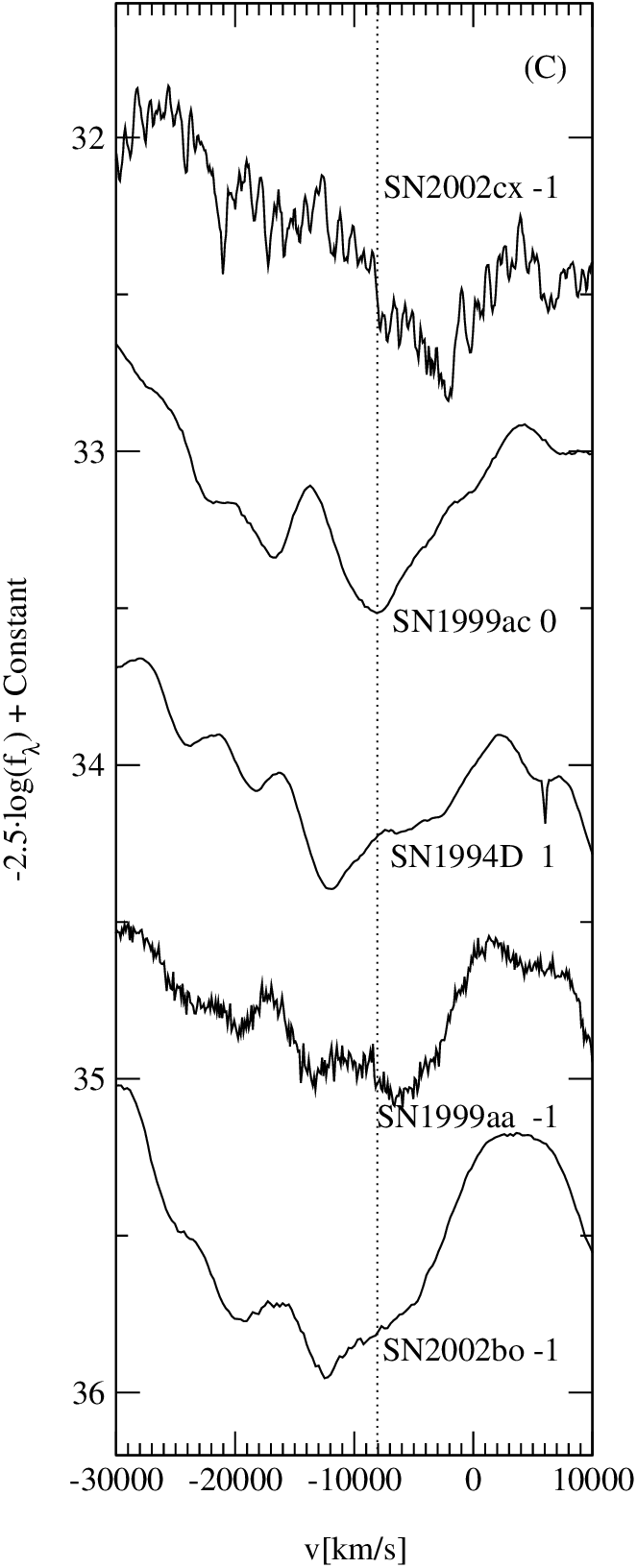}
\includegraphics[width=4cm]{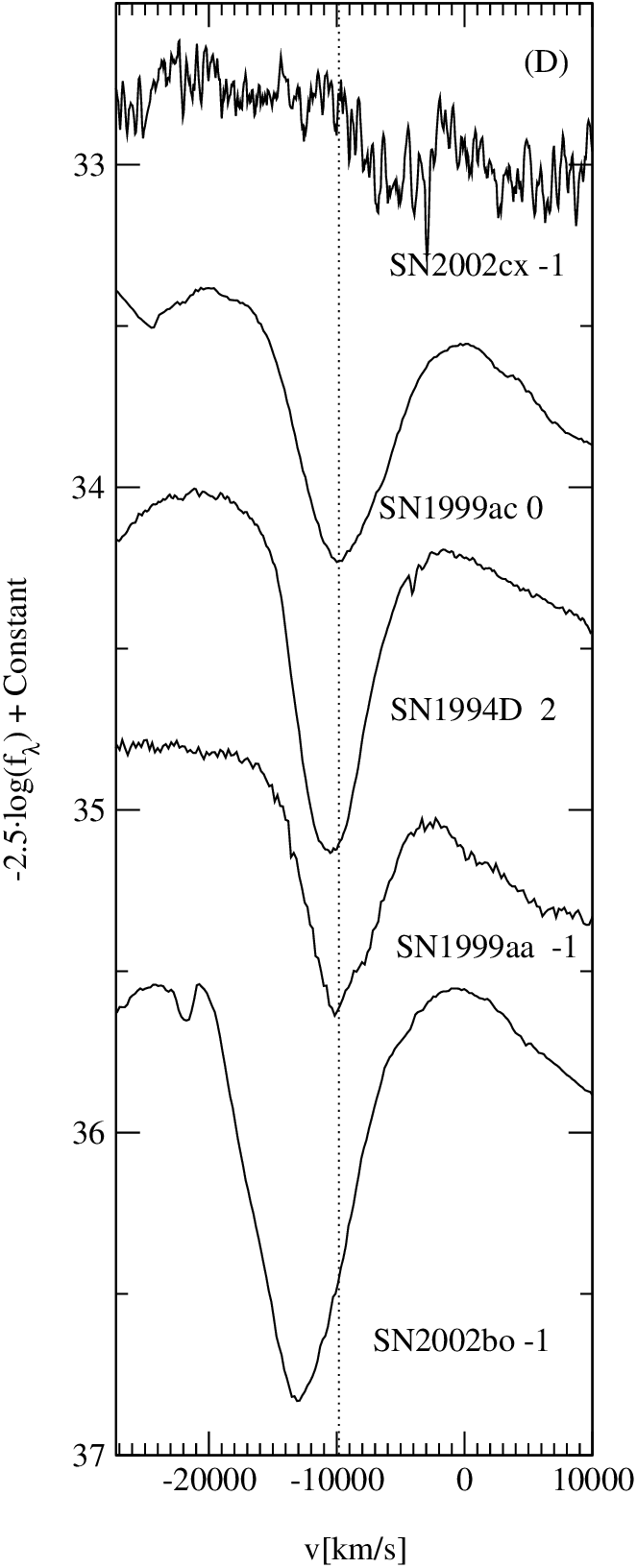}
\caption{{\it Panel A}: The spectrum at maximum light of SN~1999ac
  together with those of SN~1999aa, SN~2002cx and SN~1994D. Epochs are
  quoted in the labels.  Line identifications are taken as in
  \citet{1999AJ....117.2709L,2001PASP..113.1178L,1999MNRAS.304...67F,1996MNRAS.278..111P,1995A&A...297..509M,1993ApJ...415..589K,1992ApJ...397..304J}
  {\it Panel B}: Comparison of the same SNe in the region of
  \ion{Ca}{2}~H\&K in velocity space. {\it Panel C}: Comparison of the
  same SNe in the region of \ion{Fe}{2}~$\lambda$5083.4 in velocity
  space. {\it Panel D}: Comparison of the same SNe in the region of
  \ion{Si}{2}~$\lambda$6355 in velocity space.  In {\it Panel B,
      C and D} vertical dotted lines indicate the position of the feature
    minima for SN~1999ac.}
  \label{comp99ac0_new}
\end{figure*}

The ejecta geometry of the same SNe at around maximum light is
compared in Figure \ref{comp99ac0_new}, {\it Panel A }. As in the
previous epoch the most noticeable differences are in the iron and
silicon absorption features. Furthermore, \ion{Ca}{2}~H\&K also
appears to have a different line profile in each supernova.

The line profile in velocity space of \ion{Ca}{2}~H\&K ({\it Panel B})
of SN~1999ac is comparable with that of SN~2002bo and has a single
minimum while SN~1994D and SN~1999aa have double minima ( see
\citet{1997ApJ...485..812N,2004AJ....128..387G} a discussion of the
double minima in these SNe.).

At this epoch
the velocity of the minima are approximately all comparable with
differences of the order of a thousand kilometer per second. The only
exception is SN~2002cx for which this feature is much weaker and at
lower velocity.

 \ion{Fe}{2}~\wl5083.4 ({\it Panel C}) is at lower velocity in SN~1999ac
than in SN~1994D, SN~2002bo and SN~1999aa, while only SN~2002cx shows
even lower values. This is consistent with the iron layer of SN~1999ac
being deeper into the atmosphere than that of the other objects as
already mentioned for the case of \ion{Fe}{3}~\wl4404.  At this epoch
\ion{Si}{2}~\wl6355 ({\it Panel D)} also shows lower values than all the
other SNe analyzed except for  SN~2002cx where this line is
 considerably weaker.

\subsubsection{Late time spectra}

\begin{figure*}
\centering
\includegraphics[width=12cm]{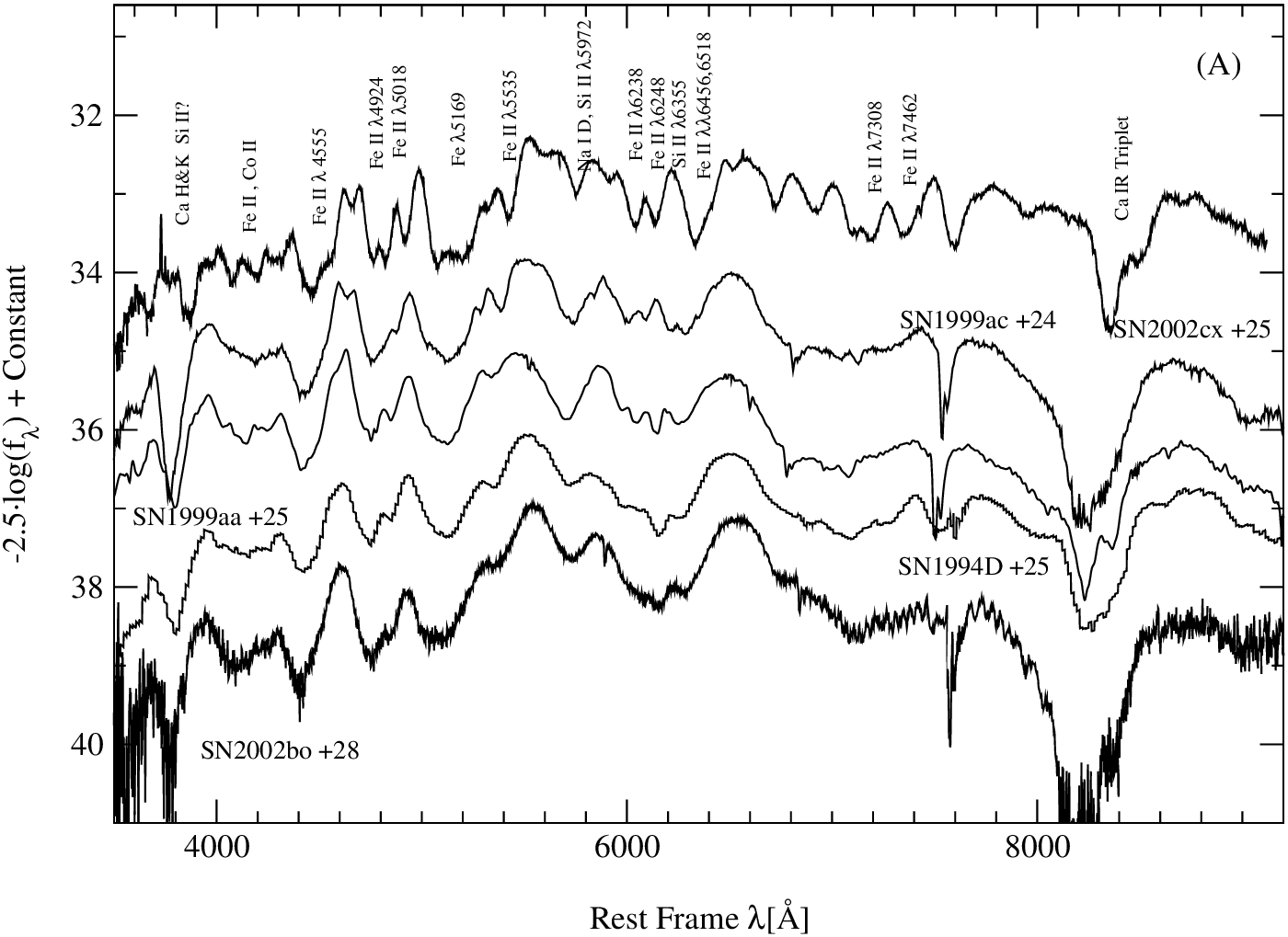}\\
\includegraphics[width=4cm]{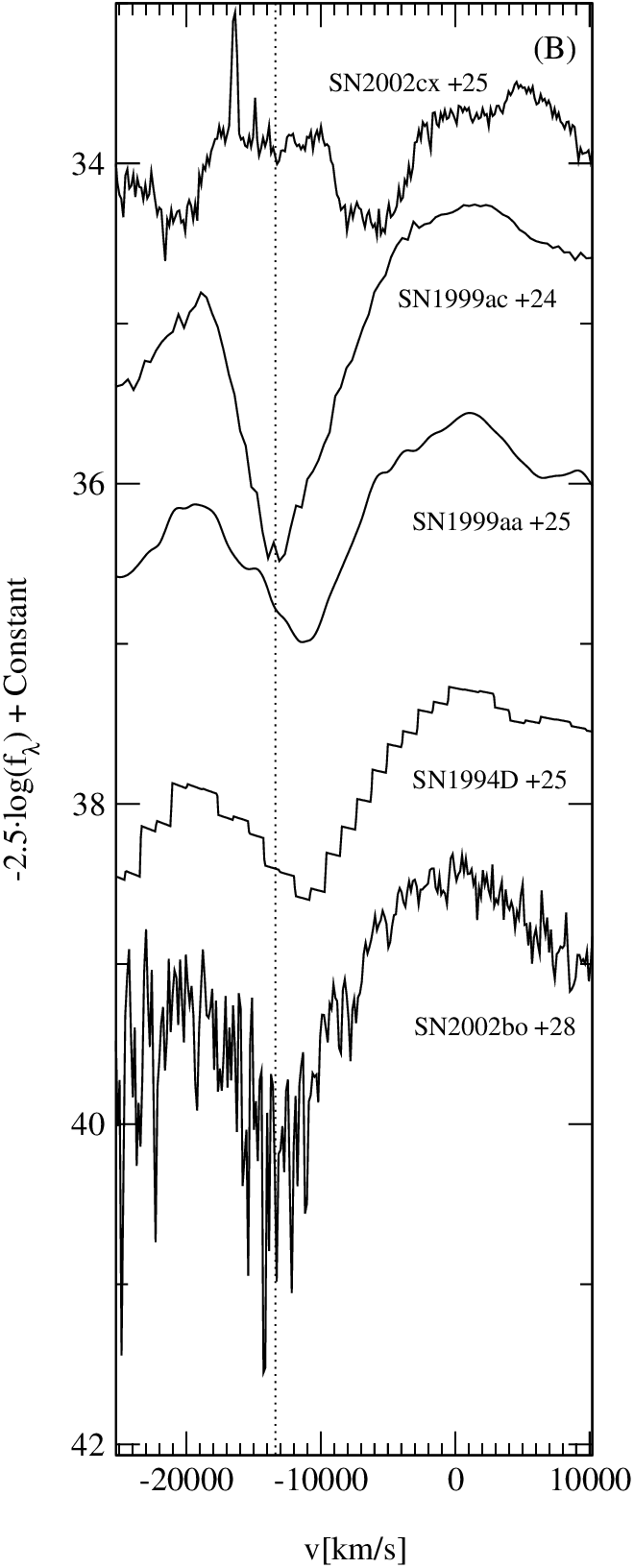}
\includegraphics[width=3.86cm]{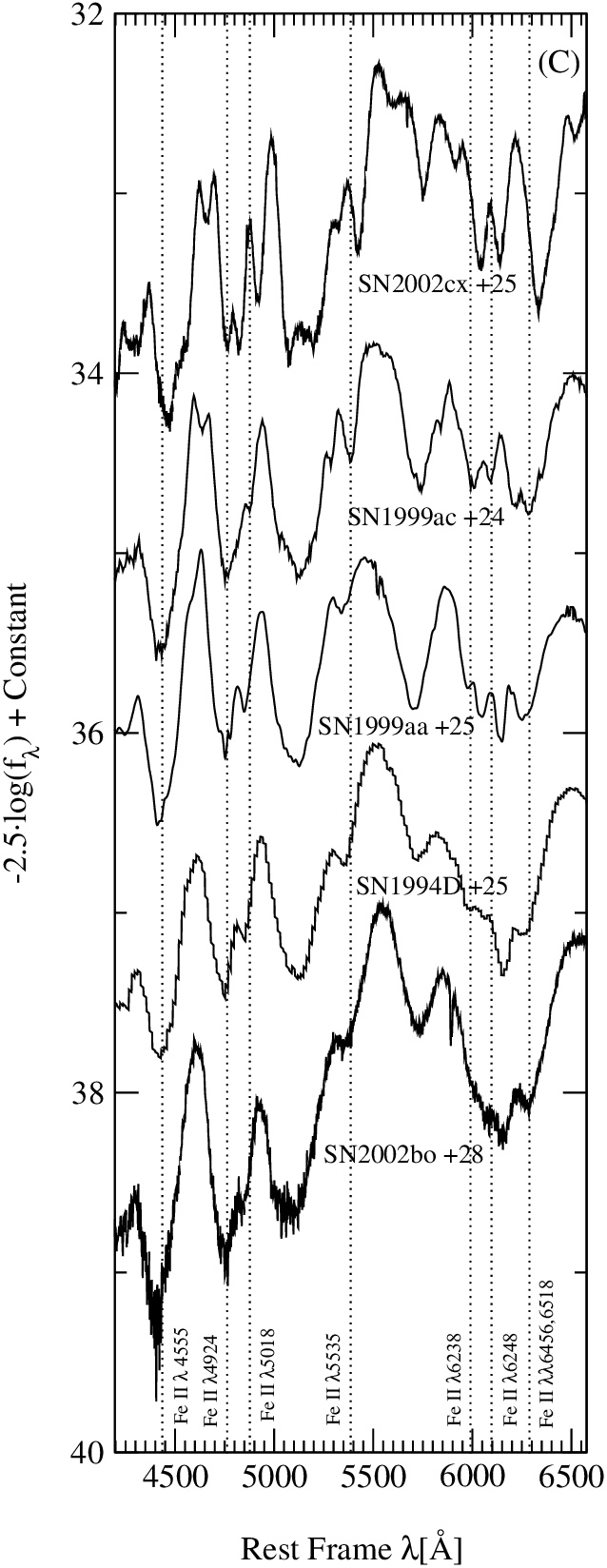}
\includegraphics[width=4cm]{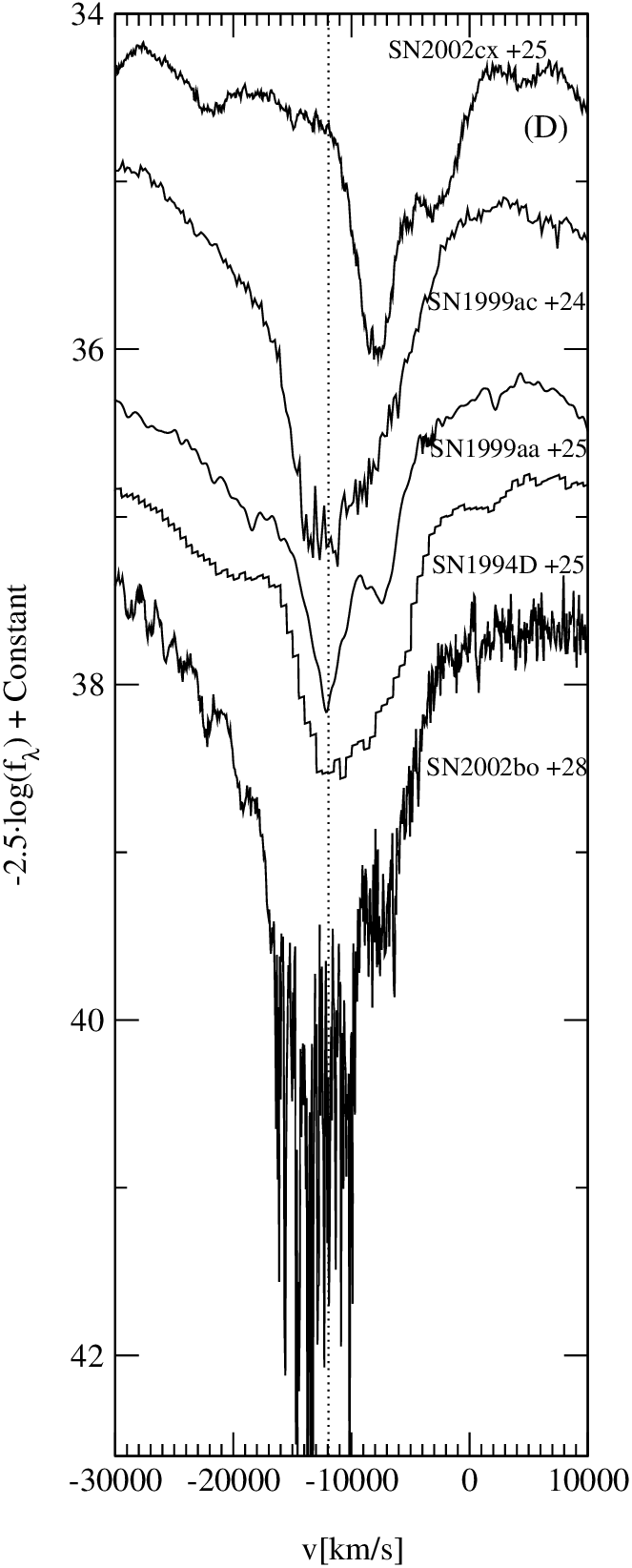}
\caption{{\it Panel A}: The $+$24 day spectrum of SN~1999ac together
  with those of SN~1999aa, SN~2002cx and SN~1994D. Epochs are quoted
  in the labels.  Line identifications are taken as in
  \citet{1999AJ....117.2709L,2001PASP..113.1178L,1999MNRAS.304...67F,1996MNRAS.278..111P,1995A&A...297..509M,1993ApJ...415..589K,1992ApJ...397..304J}
  {\it Panel B}: Comparison of the same SNe in the region of
  \ion{Ca}{2}~H\&K in velocity space. {\it Panel C}: Comparison of the
  same SNe in the \ion{Fe}{2} lines region in wavelength space. {\it
  Panel D}: Comparison of the same SNe in the region of \ion{Ca}{2}~IR
  triplet in velocity space. In {\it Panel B,
      C and D} vertical dotted lines indicate the position of the feature
    minima for SN~1999ac. }

  \label{comp99ac24_new}
\end{figure*}

In late time spectra, showed in {\it Panel A} of Figure
\ref{comp99ac24_new}, the absorption features are mainly formed by
iron lines. The velocity distribution and relative abundances of iron
in the SNe analyzed can be discerned by inspection of the different
strengths of the small absorptions and peaks in the spectra.  In {\it
  Panel C}, vertical dotted lines mark the position of the various
iron line minima. SN~1999ac shows lower velocity compared to all SNe
with the usual exception of SN~2002cx. \ion{Si}{2} at this epoch is
too weak and blended with iron lines to disentangle its complete line
profile and is not analyzed further.  \ion{Ca}{2}~H\&K ({\it Panel B})
is still strong, and for SN~1999ac this absorption feature has the
highest velocity among the analyzed SNe; The \ion{Ca}{2}~IR, {\it
  Panel D}, has a velocity similar to the other analyzed SNe.

\subsection{Velocity time evolution}
\label{cavel}
\begin{figure*}
\centering\includegraphics[width=16cm]{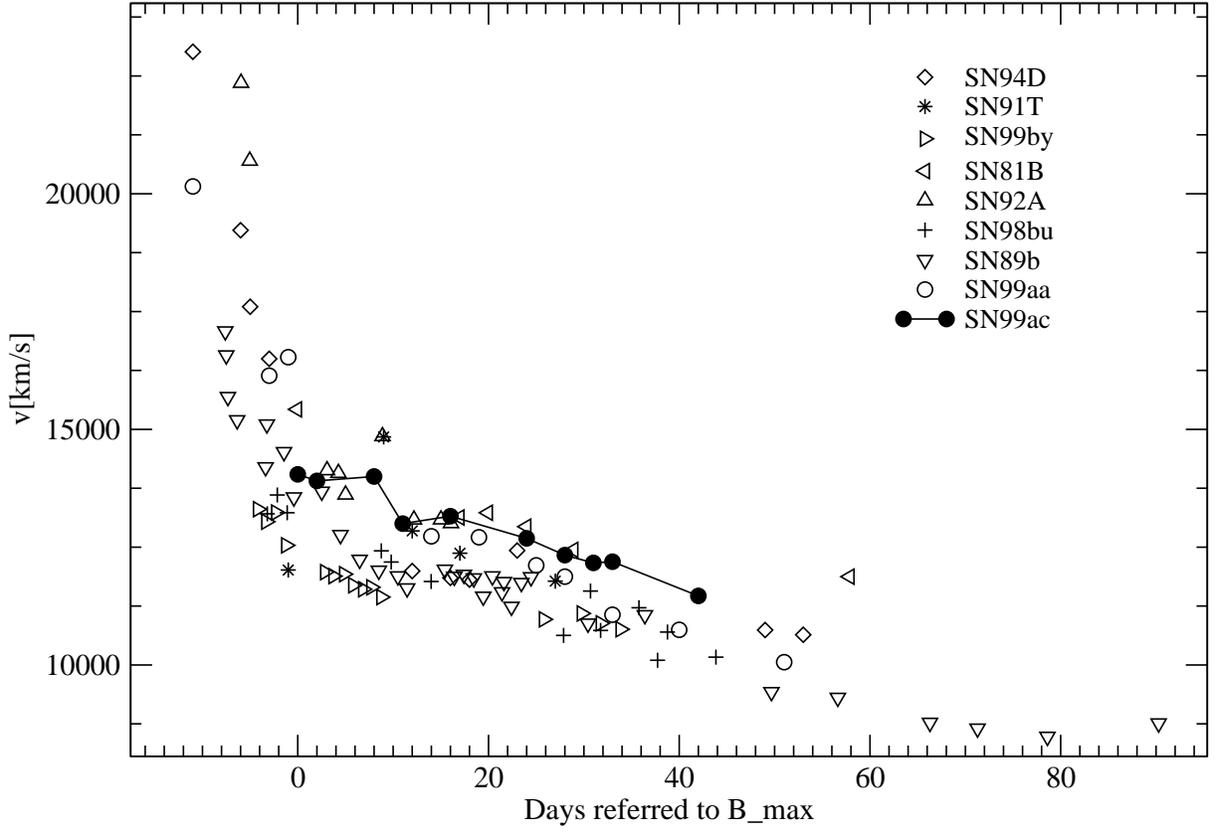}
  \caption{Expansion velocities  as measured from the
  minima of Ca~{\sc ii} H\&K of SN~1999ac compared with the values of other SNe
  taken from
  \citet{1994AJ....108.2233W,2004ApJ...613.1120G,1993ApJ...415..589K,1996MNRAS.278..111P,1999ApJS..125...73J}
  and references therein. Values for SN~1999ac are marked as filled
  circles. Measured values are reported in Table \ref{tabledata}
  together with those measured for SN~1999aa.}
  \label{CaHK_vel}
\end{figure*}

The most striking peculiarity of SN~1999ac seems to be the
concomitance of high velocity calcium with low velocity iron and
silicon. This characteristic becomes more evident with time.  Table
\ref{tabledata} lists the velocities inferred from the minimum of the
\ion{Ca}{2}~H\&K feature, together with those measured for SN~1999aa
\citep{2004AJ....128..387G}.  Figure \ref{CaHK_vel} shows these
velocities as a function of time compared with those of other well
observed SNe (see \citet{2004AJ....128..387G} for details on the
measurement techniques). The statistical uncertainty of the
measurements is negligible compared with the estimated 100
\kmps\ uncertainty in host galaxy recession velocity.  Spectra
obtained before maximum light do not include this feature, so at those
phases no measurement can be reported. After maximum, the \ion{Ca}{2}
H\&K velocity decreases monotonically. This is in agreement with the
trend measured for most of other SNe, but with SN~1999ac
systematically offset to higher velocity.

\begin{figure*}
\centering\includegraphics[width=16cm]{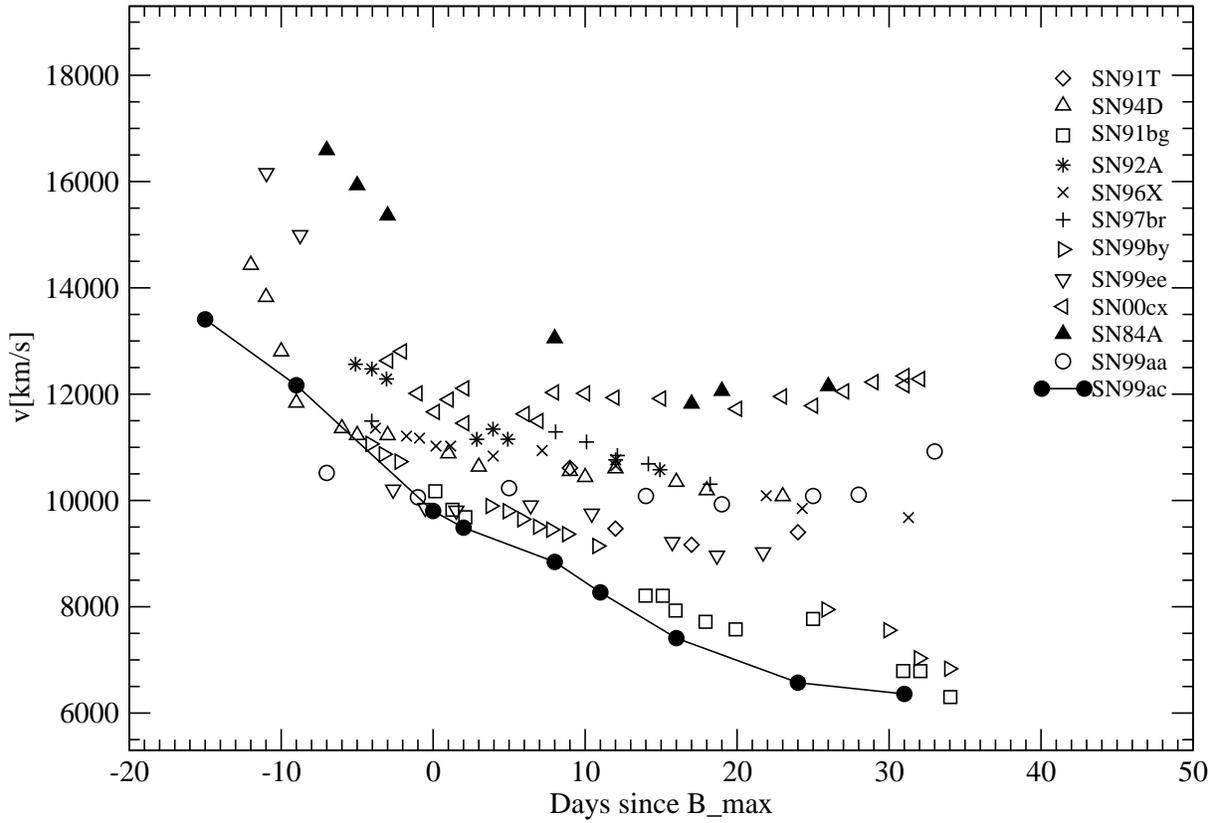}
  \caption{Expansion velocities as measured from the minima of Si~{\sc
      ii}~$\lambda$6355 of SN~1999ac compared with the values of other
    SNe taken from
    \citet{1999AJ....117.2709L,2001PASP..113.1178L,2004ApJ...613.1120G,2001MNRAS.321..254S}
    and references therein. Values for SN~1999ac are marked as filled
    circles. Measured values are reported in Table \ref{tabledata}
    together with those measured for SN~1999aa.}
  \label{siII_vel}
\end{figure*}

\begin{figure*}
\centering\includegraphics[width=8cm]{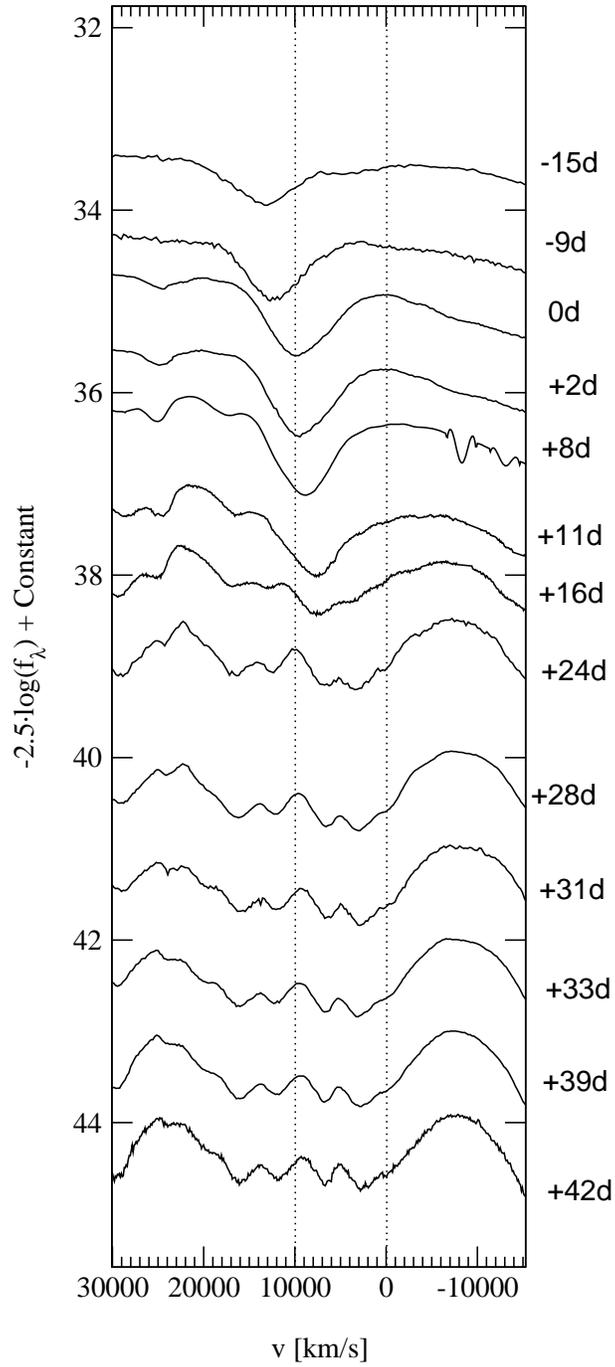}
  \caption{Evolution of the \ion{Si}{2} \wl 6355 feature in Doppler
  space.  The velocity at the minimum of the absorption feature is at
  13,000 \kmps\ at $-15$ days, and moves to 8,000 \kmps\ at $+11$ days
  relative to maximum.  By day +16, contamination by \ion{Fe}{2} lines
  has clearly begun.}
  \label{siII_doppler1}
\end{figure*}

The evolution of the velocity inferred from the minimum of \ion{Si}{2}
\wl 6355 in Figure \ref{siII_vel} is even more peculiar.  (The
measured values are reported in Table \ref{tabledata} together with
those measured for SN~1999aa.)  The \ion{Si}{2} velocity of SN~1999ac
appears to be monotonically decreasing with time as is usually the
case for dimmer supernovae, like SN~1991bg or SN~1999by. However,
SN~1999ac was not intrinsically dim (M$^{max}_{B}$=$-18.98$($39$)
\citet{2003PASP..115..453L}). Some events, such as SN~2000cx or
SN~1999aa, maintain a rather constant \ion{Si}{2} velocity as a
function of time. It should be noted that after day $+11$ the
measurements may become less meaningful because of possible blending
with \ion{Fe}{2}.  Figure \ref{siII_doppler1} illustrates this
ambiguity clearly: By day $+24$, the entire feature can no longer be
assumed to arise from \ion{Si}{2} \wl 6355 alone, since the peak of
the emission feature has shifted to the red of the rest wavelength
(zero velocity).  Still, it is clear that the velocity evolution of
the feature out to day $+11$ is extreme among SNe Ia.

\section{Type Ia SN parameter space.}
\label{parameter}
Type Ia supernovae are currently believed to be a multiparameter class
of objects. The standard paradigm is, however, to describe the
intrinsic spread of properties through one parameter: the light curve
width stretch, {\it s}, or equivalently, through the light curve
decline rate \dmft. This, indeed has proved to be successful in
reducing the intrinsic spread in brightness and in enabling the use of
Type Ia SNe as high-quality distance indicators. Several other
parameters have been proposed with the goal of fully describing SNe~Ia
and their parameter space. In the following sections we compare the
measurements of the parameters \rsi\ \citep{1995ApJ...455L.147N},
\vten\ \citep{1993AJ....105.2231B} and
\vdot\ \citep{2004astro.ph.11059B} for SN~1999ac with those of the
dataset presented in \citet{2004astro.ph.11059B}. For completeness we
report here also the data for SN~1999aa analyzed in
\citet{2004AJ....128..387G}.  The measured values for both supernovae
are reported in Table \ref{tabledata1}. The symbols used in the
following figures are chosen to be as those in
\citet{2004astro.ph.11059B} to maintain the same distinction among the
three clusters found in their analysis.

\subsection{\dmft\ versus \rsi}

\begin{figure*}
\centering\includegraphics[width=16cm]{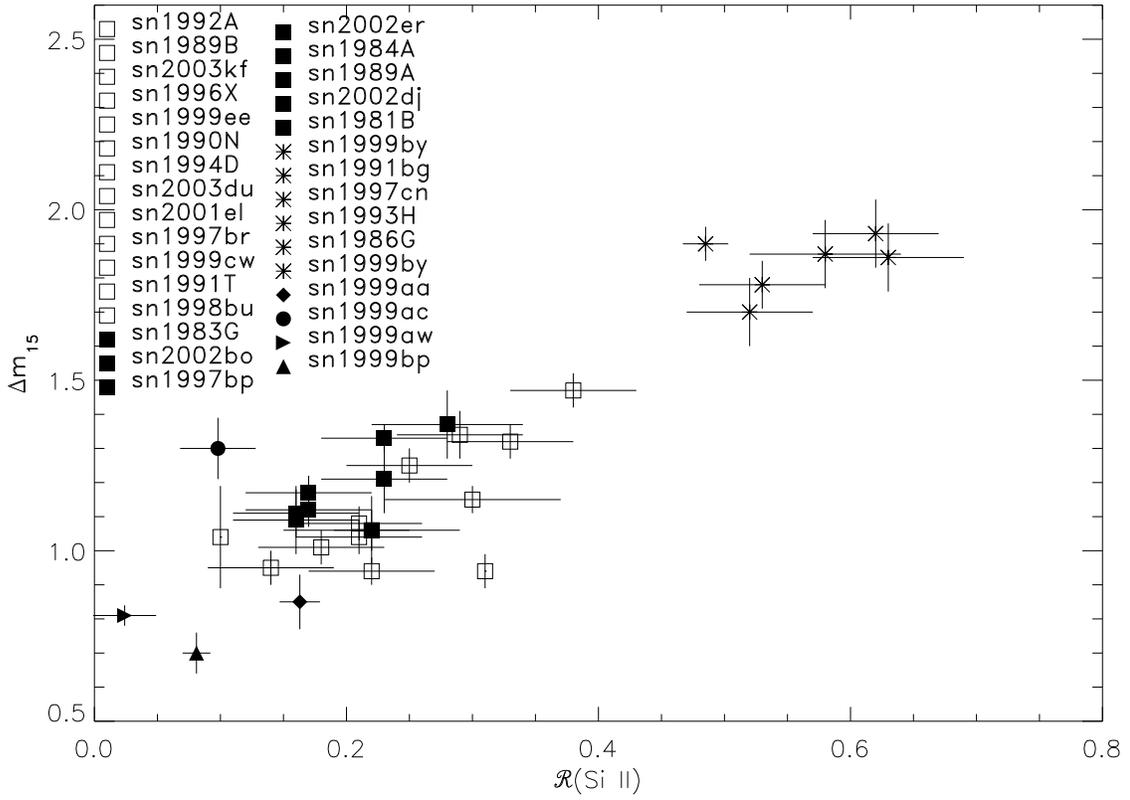}
  \caption{Light curve decline rate \dmft\ versus \rsi\ for the SNe Ia
    in \citet{2004astro.ph.11059B}. SN~1999ac generally supports the
    trend toward higher luminosity at lower \rsi\ but with a smaller
    \rsi\ with respect to normal SNe. Measured values are reported in
    Table \ref{tabledata1} together with those mesured for SN~1999aa.}
  \label{siratio}
\end{figure*}

Figure \ref{siratio} is a plot of \dmft\ as a function of the quantity
\rsi\ for many SNe Ia, including SN~1999ac.
\citet{1995ApJ...455L.147N} defined \rsi\ as the ratio of the depth of
the \ion{Si}{2}~5800~\AA\ absorption to that of the \ion{Si}{2}~6100
\AA\ absorption.  The authors theorize that the observed correlation
between \dmft\ and \rsi\ is driven by temperature (and hence nickel
mass).  Thus, hotter and brighter events tend to be characterized by a
small \rsi\ value; cooler and dimmer events are characterized by a
larger \rsi\ value.  The plot shows that SN~1999ac is above the
general trend. The small value of \rsi\ of SN~1999ac is consistent
with the presence of strong \ion{Fe}{3} lines in early epochs and a
high temperature of the envelope. The photometric peculiarities of
SN~1999ac, reported in section \ref{sec_intro}, suggest that for this
object \dmft\ might not be a good indicator of luminosity.  The
parameter space of this diagram is further expanded by the inclusion
of the values for SN~1999aw, SN~1999bp and SN~1999aa
\citep{folatellithesis} to which the correlation appears to apply.

\subsection{\dmft\ versus \vten}

\begin{figure*}
\centering\includegraphics[width=16cm]{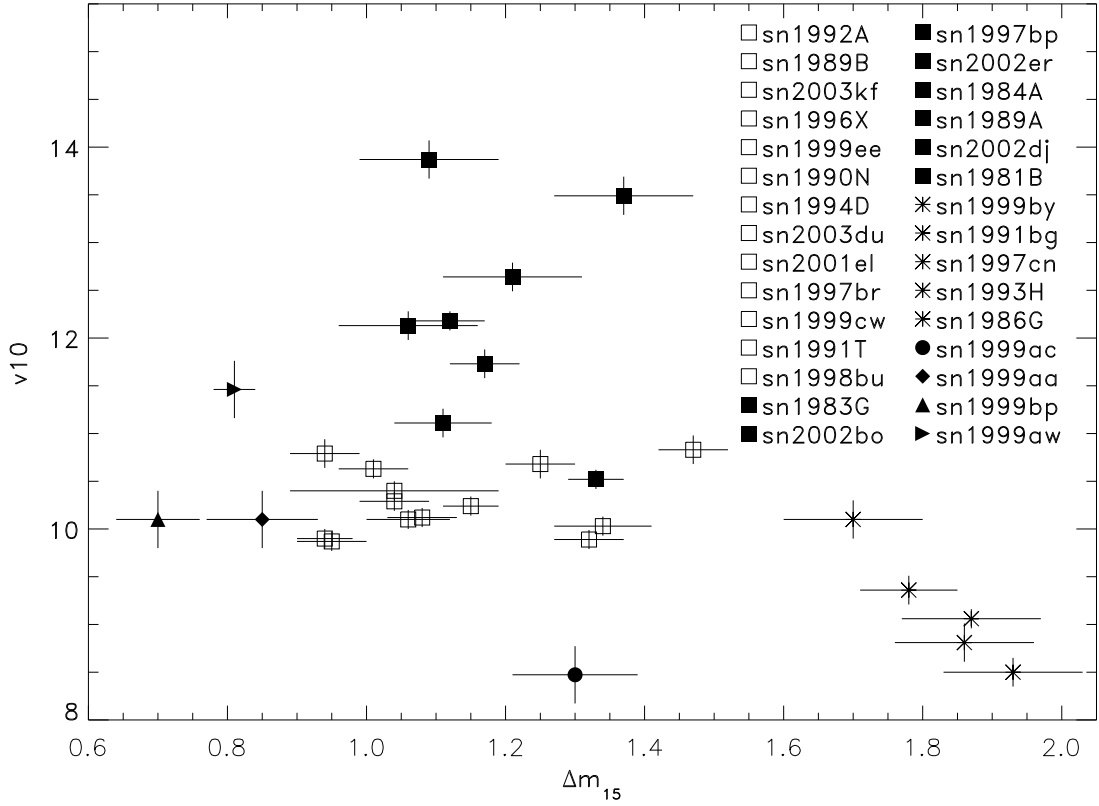}
  \caption{Plot of \dmft\ versus \vten.  SN 1999ac falls in a
  relatively unpopulated region of the plot with low value of \vten.
  Measured values are reported in Table \ref{tabledata1} together with
  those measured for SN~1999aa.}
  \label{dm15vsv10}
\end{figure*}

\citet{2000ApJ...543L..49H} investigated the spectroscopic diversity
of SNe Ia by plotting values of \rsi\ against \vten\ (the blueshift
measured in the \ion{Si}{2} \wl 6355 feature at ten days after maximum
light).  The authors reasoned that if SNe Ia were a one-dimensional
family based on the mass of synthesized $^{56}$Ni, then the two
observables would be correlated.  Unable to discern such a
correlation, they reasoned that the differences reflect differences in
the explosion mechanism itself.  In Figure \ref{dm15vsv10}, we plot
\dmft\ (a proxy for \rsi) against \vten, including observed values
from SN~1999ac, SN~1999aw, SN~1999bp and SN~1999aa.  SN~1999ac
possesses a low value for \vten\, but a normal value of
\dmft\ \citep{2003PASP..115..453L}, similar to that of SNe 1989B,
1994D, and 1996X, making SN~1999ac unique in falling outside the
apparent groupings. The same would be true plotting \rsi\ instead of
its proxy \dmft. While SN 1999ac is spectroscopically similar to SNe
1990N and 1999aa (except with lower velocities), its placement on the
diagram away from those events indicates the possibility of even
higher dimensions of diversity in these objects.

The parameter space
of this diagram is further expanded by the inclusion of the values for
SN~1999aw, SN~1999bp and SN~1999aa, though perhaps in directions that
appear more consistent with the trends among the groupings.

\subsection{\dmft\ versus \vdot}

\begin{figure*}
\centering\includegraphics[width=16cm]{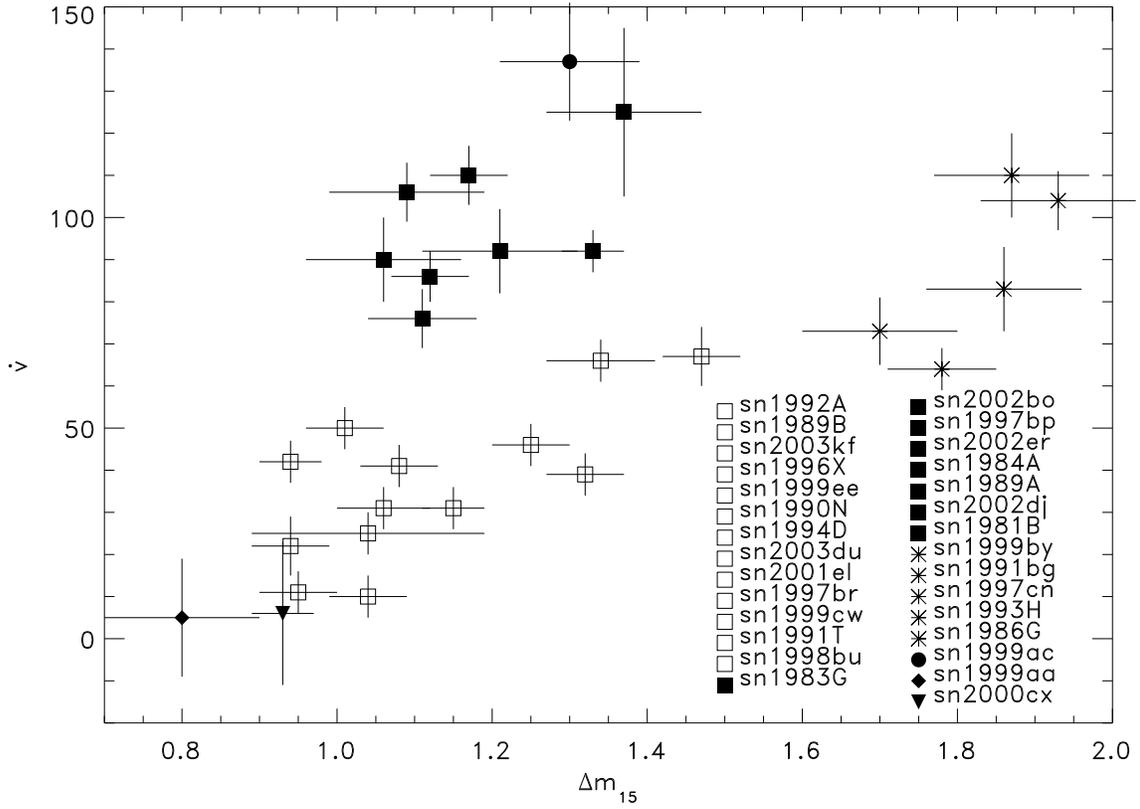}
  \caption{Plot of \dmft\ versus \vdot.  SN 1999ac generally falls in
  the region of the plot where other normal SNe lie but has the
  highest measured \vdot.  Measured values are reported in Table
  \ref{tabledata1} together with those measured for SN~1999aa.}
  \label{dm15vsvdot}
\end{figure*}

The parameter \vdot\ was introduced in \citet{2004astro.ph.11059B} as
an estimate of the expansion velocity time derivative computed after
B-band maximum light.  The authors found a weak correlation with
\dmft, Fast declining, under-luminous supernovae  show large
\vdot. Slow declining supernovae show small \vdot, while normal
supernovae can have both large and small values.  In the \dmft\
versus \vdot\ plane (Figure \ref{dm15vsvdot}) SN~1999ac falls on the high edge of the normal
supernovae with the highest \vdot\ measured to date -- similar to that of
SN~1983G, which had different \vten\ and \rsi. The parameter space of
this diagram is again expanded by the inclusion of the values for
SN~1999aa and SN~2000cx which are found to have very small values of
\vdot\ \citep{2004AJ....128..387G}.

\subsection{\vdot\ versus \vten}

\begin{figure*}
\centering\includegraphics[width=16cm]{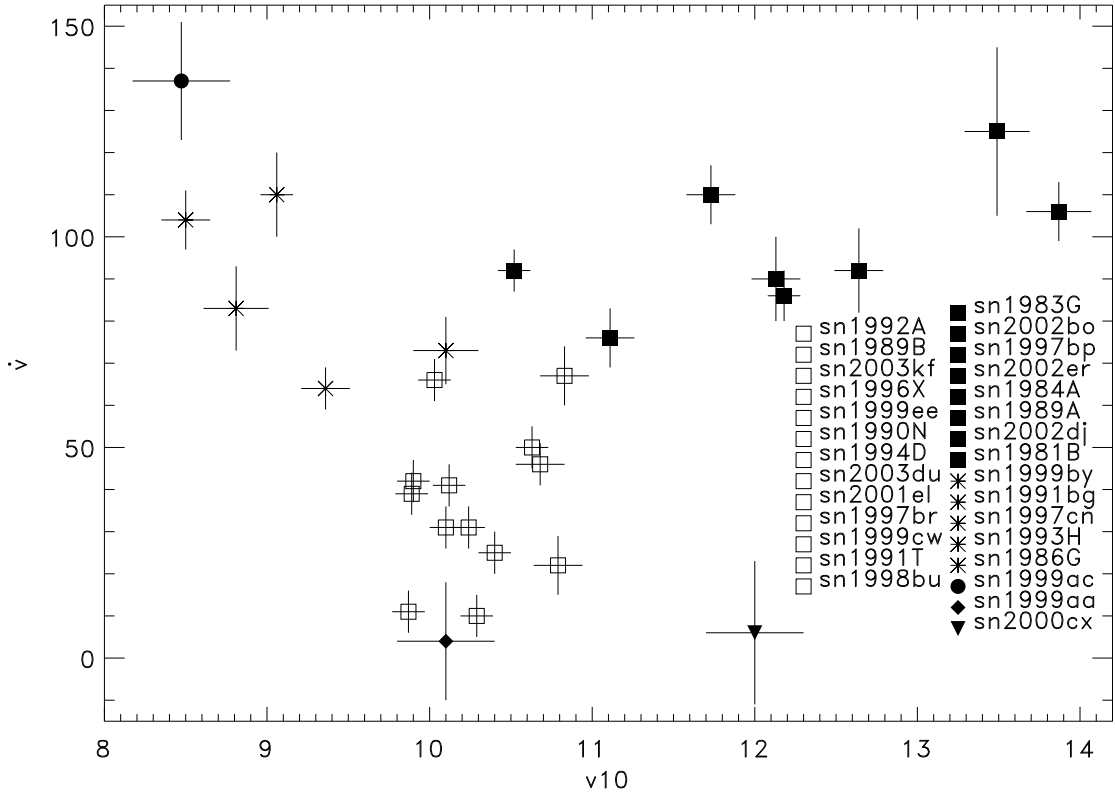}
  \caption{Plot of \vdot\ versus \vten.  SN~1999ac falls in a
  relatively unpopulated region of the plot with the highest \vdot\
  and lowest \vten. SN~2000cx falls in a unpopulated region of the
  plot with low \vdot\ and high \vten. Measured values are reported in
  Table \ref{tabledata1} together with those measured for SN~1999aa.}
  \label{vdotvsv10}
\end{figure*}

The plot of \vdot\ versus \vten\, shown in Figure \ref{vdotvsv10},
was left unexplored in \citet{2004astro.ph.11059B} but it is
interesting to note that the three SN groups identified by the authors
still populate different regions of this plot, meeting each other at
intermediate values.  On this plot underluminous supernova tend to
have small values of \vten\ and high values of \vdot. Normal
supernovae populate the central part of the plane while supernovae with
fast expansion velocity (i.e. high \vten) have also high \vdot.
SN~1999ac has the highest \vdot\ and the lowest \vten\ among the
supernova measured, making it similar to the faint supernovae but on
the extreme end. However, as noted, SN~1999ac did not appear to be
dim. It also should be noted  that SN~2000cx falls in an
unpopulated region of the plot with high \vten\ and low \vdot.

   \section{Conclusions}

\label{sec_conclusion}
We have presented spectroscopic observations of SN~1999ac from $-15$ to
$+42$ days with respect to B-band maximum light.  The earliest spectra
are similar to those of SN~1999aa, but share some characteristics of
spectroscopically normal SNe like SN~1990N.  Notable is the early 
appearance of iron features in the spectrum at $+11$ days
and the lack of any real evolution in the features after $+24$ days.

Using synthetic spectra, we have unambiguously identified the presence
of \ion{C}{2} at high velocity in the earliest spectrum at $-15$ days.
However, just six days later, all significant traces of \ion{C}{2} have
disappeared.  This indicates that obtaining spectra even earlier than
$-10$ days with respect to maximum may be required to reliably probe
the outer layers of SNe Ia.  Obtaining this level of efficiency from
an observing program requires careful coordination, similar to that
achieved by the European Research Training
Network\footnote{http://www.mpa-garching.mpg.de/~rtn/} search for SNe Ia,
and the Nearby Supernova Factory \citep{2002SPIE.4836...61A}.

Comparing the spectra of SN~1999ac with those of other supernovae we
find an unusual ejecta geometry. Iron and silicon lines
appear to be formed in deeper atmosphere layers than the corresponding
lines for other supernovae with the exception of the extreme case of
SN~2002cx. Calcium lines, however, are found to be formed in high
velocity layers. The same trends are confirmed when analyzing the time
evolution of velocities as derived from the minimum of \ion{Ca}{2}
H\&K and \ion{Si}{2} \wl 6355.  The former shows a trend consistent
with normal SNe Ia, though the values are slightly higher than
average, while the latter shows monotonically decreasing
values that follow the trend of under-luminous SNe Ia.

We have presented measurements for spectroscopically derived
observables such as \rsi, \vten and \vdot.  While the values
for SN~1999ac weakly support the correlation between \rsi\ and
\dmft, the position of SN~1999ac on the \dmft\ versus \vten\ plane
is somewhat off the main trend. In the plane \vdot\ versus \dmft\
SN~1999ac falls in a region of normal supernovae but has the highest
value of \vdot, while in the plane \vdot\ versus \vten\ it appears to
be on the extreme of faint supernovae with the highest \vdot\ and
lowest \vten. However, SN~1999ac was not reported as a dim supernova.
This analysis points out that SN~1999ac is unlike any other known
supernova.

   \acknowledgements

The authors thank David Branch and Adam Fisher for
providing the SYNOW code.  The research presented in this article made
use of the SUSPECT\footnote{http://www.nhn.ou.edu/$\sim$suspect}
Online Supernova Spectrum archive, and the atomic line list of
\citet{1993KurCD...1.....K}.  This work is based on observations made
with: the Nordic Optical Telescope, operated on the island of La Palma
jointly by Denmark, Finland, Iceland, Norway, and Sweden, in the
Spanish Observatorio del Roque de los Muchachos of the Instituto de
Astrofisica de Canarias; the Apache Point Observatory 3.5-meter
telescope, which is owned and operated by the Astrophysical Research
Consortium; the Lick Observatory Shane 3.0-m Telescope; the Cerro
Tololo Inter-American Observatory 4-m Blanco Telescope; the European
Southern Observatory 3.6m telescope (program ID: 63.0-0347(A)) and the
Kitt Peak National Observatory Mayall 4-m Telescope. We wish to thank
the staff of these observatories for their assistance in obtaining the
data presented herein. This work was supported in part by "The Royal
Swedish Academy of Sciences".  G. Garavini acknowledges support from
the Physics Division, E.O. Lawrence Berkeley National Laboratory of
the U.S. Department of Energy under Contract No.  DE-AC03-76SF000098.
A. Mour\~ao acknowledges financial support from Funda\c{c}\~ao para a
Ci\^encia e Tecnologia (FCT), Portugal, through project
PESO/P/PRO/15139/99; S. Fabbro thanks the fellowship grant provided by
FCT through project POCTI/FNU/43749/2001.

\clearpage 
\clearpage 

\clearpage

\clearpage

\clearpage 

\clearpage 

\clearpage 

\clearpage 
\clearpage 

\clearpage 

\clearpage 

\clearpage 

\clearpage

\clearpage

\clearpage

\clearpage 

\clearpage

\begin{table*}
\caption{Data set specifications.\label{tabdata}}
\begin{center}
\begin{tabular}{lrllcllr}
\tableline\tableline
JD & Epoch\tablenotemark{f} & Telescope &Instrument&$\lambda$ Range\tablenotemark{g}&
    FWHM$\tablenotemark{a,g}$&$\langle S/N \rangle$$\tablenotemark{b}$& 
Comments\\ 
-2400000   & ref $B_{max}$ & &  &[\AA] &[\AA] \\
\tableline
51236.89 &  -15 & APO     &  DIS & 3703-10307 & 11 & 125&5696.16 $\tablenotemark{c}$ \\
51240.95 &  -9  & MDM 2.4m& MARK III &3827-8860 & 16 &  63& $\tablenotemark{e}$\\
51251.39 &  +0  & ESO 3.6m    &EFOSC2 &3331-7495 & 32 & 325&$\tablenotemark{d}$ \\
51253.84 &   +2  & CTIO 4m  &RCSP &  3235-9263 & 5 & 76& $\tablenotemark{d}$\\
51253.72 &   +2  & NOT     & ALFOSC &  3285-9655 & 34 &  176&5852.06$\tablenotemark{c}$\\
51259.88 &   +8  & CTIO 4m  &RCSP & 3227-9254 & 6 & 94 &$\tablenotemark{d}$\\
51262.89 &   +11  & CTIO 4m  &RCSP & 3254-9278 & 6 &  114&$\tablenotemark{d}$ \\
51267.87 &   +16  & CTIO 4m  &RCSP &  3239-9241 & 6 &  90 & $\tablenotemark{d}$\\
51275.98 &   +24  & KPNO 4m  &T2KB &  3029-10401 & 11 &  59 &$\tablenotemark{d}$ \\
51279.85 &   +28  & ESO 3.6m   &EFOSC2 & 3341-10255 & 32 &  
237&7440.45$\tablenotemark{c,d}$\\
51282.90 &  +31  & Lick 3m  & KAST&3321-10483 & 5 & 46&5489.40$\tablenotemark{c}$\\
51284.84 &   +33  & ESO 3.6m   &EFOSC2 & 3392-10128 & 32 & 
195&7363.23$\tablenotemark{c,d}$\\
51290.84 &   +39  & ESO 3.6m   &EFOSC2 & 3344-10194 & 32 & 
146&7435.50$\tablenotemark{c,d}$\\
51293.97 &  +42 & Lick 3m  & KAST&3268-8002 & 5 &  49&5417.33$\tablenotemark{c}$\\
\tableline
\end{tabular}
\tablenotetext{a}{Average over wavelength.}
\tablenotetext{b}{Average signal-to-noise ratio per wavelength bin.}
\tablenotetext{c}{Beginning of red channel, [\AA].}
\tablenotetext{d}{Negligible 2$^{nd}$ order contamination.}
\tablenotetext{e}{Possible 2$^{nd}$ order contamination above 7500~\AA.}
\tablenotetext{f}{Rest Frame days.}
\tablenotetext{g}{Rest Frame wavelength.}
\end{center}
\end{table*}

\clearpage

\begin{table*}
\caption{Synow parameters for  $-$15 days. The fit is shown in Figure \ref{synow-15}. 
v$_{phot}$=13000~km s$^{-1}$,
$T_{bb}=11200$ K.\label{table-15}}
\begin{center}
\begin{tabular}{llllll}
\tableline\tableline
Ion  & $\tau$ & $v_{min}$&$v_{max}$&$T_{exc}$ &$v_{e}$\\
  &&$10^3$~km$s^{-1}$&$10^3$~km$s^{-1}$&\small$10^{3}$K &$10^3$~km$s^{-1}$\\
\tableline
C~{\sc ii}&0.038&16&40&15&5\\
C~{\sc iii}&0.2&-&14.2&15&5\\
O~{\sc i}&0.2&-&40&15&5\\
Mg~{\sc ii}&0.15&-&40&15&5\\
Si~{\sc ii}&0.65&14.2&40&15&5\\
Si~{\sc iii}&0.42&-&17&15&5\\
S~{\sc ii}&0.2&-&17&15&5\\
Ca~{\sc ii}&1.5&16&40&15&5\\
Fe~{\sc iii}&0.55&14.5&18&12&5\\
Co~{\sc ii}&0.006&-&40&15&5\\
Ni~{\sc iii}&5&-&40&12&5\\
\hline
\end{tabular}

\end{center}
\end{table*}

\clearpage

\begin{table*}
\caption{Synow parameters for $-$9 days. The fit is shown in Figure 
\ref{synow-15}.v$_{phot}$=11800~km s$^{-1}$,
$T_{bb}=13800$ K.\label{table-9}}
\begin{center}
\begin{tabular}{llllll}
\tableline\tableline
Ion  & $\tau$ & $v_{min}$&$v_{max}$&$T_{exc}$ &$v_{e}$\\
  &&$10^3$~km$s^{-1}$&$10^3$~km$s^{-1}$&\small$10^{3}$K &$10^3$~km$s^{-1}$\\
\tableline
C~{\sc ii}&0.015&16&40&12&5\\
C~{\sc iii}&0.75&-&12.8&12&5\\
O~{\sc i}&0.1&-&40&12&5\\
Mg~{\sc ii}&0.2&-&40&12&5\\
Si~{\sc ii}&1.2&-&40&12&5\\
Si~{\sc iii}&0.6&-&16&12&5\\
S~{\sc ii}&0.2&-&17&12&5\\
Ca~{\sc ii}&1.5&16&40&12&5\\
Fe~{\sc iii}&0.65&-&18&12&5\\
Ni~{\sc iii}&7.0&-&40&12&5\\
Co~{\sc ii}&0.045&-&40&12&5\\
\hline
\end{tabular}
\end{center}
\end{table*}

\clearpage

\begin{table*}
\caption{Measurements of the expansion velocity inferred from
\ion{Ca}{2}~H\&K and \ion{Si}{2}~$\lambda$6355 for SN~1999ac and
SN~1999aa \citep{2004AJ....128..387G}. The uncertainties on the velocity
measurements are assumed to be 100 \kmps, for further information see \S\ref{cavel}. 
\label{tabledata}}
\begin{center}
\begin{tabular}{ccc|ccc}
\tableline\tableline
&SN~1999ac&&&SN~1999aa&\\
\hline
Epoch&\ion{Ca}{2}~H\&K&\ion{Si}{2}&Epoch&\ion{Ca}{2}~H\&K&\ion{Si}{2}\\
 days&[\kmps]&[\kmps]&days&[\kmps]&[\kmps]\\
\hline
-15&    -    & 13409   & -11 & 20155   &   -   \\
-9 &    -    & 12169   & -7  &  -      & 10519 \\
0  & 14045   & 9801    & -3  & 16138   &   -   \\
2  & 13907   & 9488    & -1  & 16531   & 10060 \\ 
8  & 14000   & 8844    & 5   &   -     & 10233 \\
11 & 12999   & 8269    & 14  & 12730   & 10083 \\ 
16 & 13156   & 7410    & 19  & 12710   &  9928 \\ 
24 & 12687   & 6572    & 25  & 12114   & 10085 \\ 
28 & 12331   & 6361    & 28  & 11874   & 10107 \\ 
31 & 12167   &   -     & 33  & 11066   & 10922 \\ 
33 & 12192   &   -     & 40  & 10747   &   -   \\
42 & 11465   &   -     & 51  & 10058   &   -   \\
\hline
\end{tabular}

\end{center}
\end{table*}

\clearpage

\begin{table*}
\caption{Measured values of M$_{B}^{max}$, \dmft, \rsi, \vten, \vdot\
for SN~1999ac and SN~1999aa. \label{tabledata1}}
\begin{center}
\begin{tabular}{ll|l}
\tableline\tableline
&SN~1999ac&SN~1999aa\\
\hline
M$_{B}^{max}$&-18.98(0.39)$\tablenotemark{a}$&-19.14(0.78)\\
\dmft&1.30(0.09)$\tablenotemark{a}$&0.85(0.08)\\
\rsi &0.098(0.030)&0.163(0.016)\\
\vten&8470(300)&10100(300)\\
\vdot&137(14)&4(14)\\

\hline
\end{tabular}
\tablenotetext{a}{As reported in \citet{2003PASP..115..453L}.}
\end{center}
\end{table*}

\end{document}